\newif\ifsingle
\ifsingle
\documentclass[11pt,draftclsnofoot, onecolumn]{IEEEtran}		
\else		
\documentclass[10pt,final, twocolumn]{IEEEtran}
\fi

\usepackage{times}
\usepackage{amsmath,dsfont}
\usepackage{amssymb,amsthm}
\usepackage{epsfig,verbatim}
\usepackage{setspace}
\usepackage{color}
\usepackage{cite}
\usepackage{epstopdf}
\usepackage{graphics}
\usepackage{accents}
\usepackage{acronym}
\usepackage[bookmarks,colorlinks]{hyperref}
\usepackage{booktabs}
\usepackage{mathtools}
\usepackage{enumitem}
\usepackage{Quant}
\usepackage{bm}
\usepackage{multirow}
\usepackage{subcaption}

 \usepackage[all=normal,paragraphs=tight,floats=normal,mathspacing=normal,wordspacing=tight,charwidths=tight,mathdisplays=normal,leading=normal]{savetrees}
\usepackage[ruled,linesnumbered,vlined]{algorithm2e}
\SetKwInput{KwData}{\textbf{Init}} 
\let\oldnl\nl% Store \nl in \oldnl
\newcommand{\nonl}{\renewcommand{\nl}{\let\nl\oldnl}}% Remove line number for one line

\newcommand{\myVec}[1]{{\boldsymbol{#1}}}
\newcommand{\myMat}[1]{{\boldsymbol{#1}}}
\newcommand{\mySet}[1]{\mathcal{#1}}

\newcommand{\Tsig}{{\tau_{\max}}}

\newcommand{\TDAmbMat}[1]{\myMat{A}_{#1}}
\newcommand{\TDAmbMatest}[1]{\hat{\myMat{A}}_{#1}}

\newcommand{\FDAmb}[1]{\myVec{A}^{\rm tf}_{#1}}
\newcommand{\FDAmbest}[1]{\hat{\myVec{A}}^{\rm tf}_{#1}}
 
\acrodef{kf}[KF]{Kalman filter} 
\acrodef{doa}[DOA]{directions of arrival} 
\acrodef{sh}[SH]{spherical harmonics} 
\acrodef{sgm}[SGM]{Score-based Generative Model} 
\acrodef{sde}[SDE]{stochastic differential equation} 
\acrodef{ode}[ODE]{ordinary differential equation} 
\acrodef{au}[AU]{Ambisonics upscaling} 
\acrodef{foa}[FOA]{first-order Ambisonics} 
\acrodef{hoa}[HOA]{high-order Ambisonics} 
\acrodef{fm}[FM]{Flow Matching} 
\acrodef{cfm}[CFM]{Conditional Flow Matching} 
\acrodef{vf}[VF]{vector field} 
\acrodef{ve}[VE]{Variance Exploding} 
\acrodef{vp}[VP]{Variance Preserving} 
\acrodef{cfg}[CFG]{classifier free guidance} 
\acrodef{ou}[OU]{Ornstein-Uhlenbeck}
\acrodef{sub-vp}[sub-VP]{sub-Variance Preserving}
\acrodef{di}[DI]{Directivity Index}
\acrodef{mae}[MAE]{Mean Absolute Error}
\acrodef{ctn}[CTN]{Conv-TasNet}
\acrodef{rt}[$T_{60}$]{Reverberation Time}
\acrodef{anova}[ANOVA]{analysis of variance}
\acrodef{pinn}[PINN]{physically informed neural network}
\acrodef{gan}[GAN]{Generative Adversarial Network}
\acrodef{gd}[GD]{Gradient Descent}
\acrodef{gru-u}[GRU-Uni]{GRU-Unidirectional}
\acrodef{gru-b}[GRU-Bi]{GRU-Bidirectional}
\acrodef{nfe}[NFE]{Number of Function Evaluations}

\acrodef{stft-sdr}[STFT-SDR]{STFT signal-to-distortion ratio} 
\acrodef{stft-si-sdr}[STFT-SI-SDR]{STFT scale invariant signal-to-distortion ratio} 
\acrodef{sh-rir}[SH-RIR]{Spherical Harmonics Room Impulse Response}
\acrodef{rir}[RIR]{Room Impulse Response}
\acrodef{drr}[DRR]{Direct-to-Reverberant-Ratio}
\acrodef{ism}[ISM]{Image Source Method}

\acrodef{tf}[TF]{time-frequency}
\acrodef{stft}[STFT]{short-time Fourier transform}
\acrodef{istft}[ISTFT]{inverse STFT}
\acrodef{ncsn}[NCSN++]{noise-conditioned score-matching network}
\acrodef{rd}[RD]{reverse diffusion}
\acrodef{ald}[ALD]{annealed Langevin dynamics}
\acrodef{mushra}[MUSHRA]{MUltiple Stimuli with Hidden Reference and Anchor}
\acrodef{pwd}[PWD]{plane wave decomposition}
\acrodef{cs}[CS]{compressed sensing}
\acrodef{ot}[OT]{Optimal Transport}
\acrodef{ema}[EMA]{Exponential Moving Average}

\newcommand{\RI}[1]{%\textcolor{blue}
{#1}}

\definecolor{NewColor}{rgb}{0,0,0}%{0.54, 0.17, 0.89}%

\IEEEoverridecommandlockouts
\title{
Generative Learning for Ambisonic Upscaling}

\author{
\IEEEauthorblockN{Amit Milstein, Nir Shlezinger, and Boaz Rafaely}
\thanks{ 
The preliminary findings of this research were submitted to the International Workshop on Acoustic Signal Enhancement (IWAENC 2026) as the paper \cite{milstein2025diffau}.  
The authors are with the ECE School Ben-Gurion University of the Negev, Be’er-Sheva, Israel (e-mail: amitmils@post.bgu.ac.il; \{nirshl; br\}@bgu.ac.il). This work was supported by the Israel Science Foundation (ISF) under grant no. 3314/25, and by the European Research Council (ERC) under the ERC starting grant nr. 101163973 (FLAIR). 
}
}

\begin{document}
%\ninept
% \raggedbottom

\maketitle
\pagestyle{plain}
\thispagestyle{plain}
 
%%%%%%%%%%%%%%%%
%%% Abstract %%%
%%%%%%%%%%%%%%%%
%
\begin{abstract}
\ac{au} aims to enhance the spatial resolution of sound fields by estimating \ac{hoa} components from low-order observations. While deep learning and model-based strategies have been considered for \ac{au}, both approaches exhibit significant performance degradation in realistic scenarios, where reverberant sound fields violate the directional sparsity inherent to discriminative mappings. In this work, we address \ac{au} as a generative task rather than  a deterministic reconstruction, expanding generative modeling to specifically target the recovery of spatial information in reverberant speech. We investigate two dominant continuous-time generative paradigms, adapting both \acl{sgm} and \acl{fm} to these complex acoustic settings. We provide an extensive numerical study comparing our methods against state-of-the-art baselines in various acoustic scenarios. Additionally, we conduct subjective listening tests to evaluate the perceived quality and spatial accuracy of the proposed generative framework across various reverberant scenarios. The studies reveal that Flow Matching consistently outperforms both its discriminative counterparts and Diffusion-based paradigms in all reverberant settings.
\end{abstract}

\acresetall

%----------------------------------------------------------------------------------------
%	INTRODUCTION
%----------------------------------------------------------------------------------------
\section{Introduction}\label{sec:intro}
Spatial audio technology aims to reproduce sound fields in a way that preserves the spatial cues perceived by human listeners, such as the direction and distance of sound sources~\cite{rumsey2012spatial}. By reconstructing the acoustic scene in three dimensions, spatial audio enables immersive auditory experiences and is widely used in applications including virtual and augmented reality, gaming, cinema production, teleconferencing, and music reproduction~\cite{hacihabiboglu2017perceptual}. Among  spatial audio representations, Ambisonics has emerged as a particularly attractive framework due to its flexibility in capturing, encoding, and rendering sound fields~\cite{zotter2019ambisonics}. Ambisonics represents the sound field using spherical harmonic basis functions, allowing the representation to remain independent of the recording and playback systems and enabling scalable spatial resolution.

In Ambisonics, the spatial resolution of the sound field is determined by the order of the spherical harmonic expansion. \Ac{foa} employs four channels corresponding to the zeroth- and first-order harmonics, which can be captured using relatively compact and affordable microphone arrays~\cite{gerzon1975design}. However, due to the limited number of spatial components, \ac{foa} provides only coarse spatial resolution, which can degrade localization accuracy and reduce the sense of immersion. Increasing the Ambisonic order to \ac{hoa} representations that include additional spherical harmonic components allows capturing finer spatial details~\cite{daniel2003further}. The drawback is that acquiring \ac{hoa} signals typically requires large and costly microphone arrays with many sensors, which restricts their practical deployment. Consequently, there is a growing interest in developing {\em \ac{au}} techniques that estimate high-order components from low-order recordings, enabling enhanced spatial resolution without requiring complex acquisition hardware.

%Spatial audio technology enhances the listener’s experience by accurately reproducing the direction and distance of sound sources in a three-dimensional space. It is commonly used in VR/AR, gaming, cinema, teleconferencing, and music to create immersive and realistic soundscapes~\cite{hacihabiboglu2017perceptual}. Among spatial audio formats, Ambisonics \cite{zotter2019ambisonics} stands out for its flexibility and scalability in capturing, encoding, and rendering sound fields. \Ac{foa}, which uses four channels, offers a practical advantage by requiring relatively simple hardware~\cite{gerzon1975design}. However, its spatial resolution is limited, leading to coarser localization and immersion. In contrast, \ac{hoa} offers significantly better spatial detail through more channels~\cite{daniel2003further}. However, capturing \ac{hoa} requires large, expensive microphone arrays, limiting its accessibility. This gap motivates the development of efficient upsampling techniques that can enhance \ac{foa}’s spatial resolution without the need for high-order acquisition hardware.

Early approaches to \ac{au} are predominantly model-based and rely on physical models of the sound field. A common strategy is to perform \ac{pwd} and reconstruct the missing high-order components by exploiting sparsity assumptions on the spatial distribution of sources. In particular, several works formulate the problem using \ac{cs}, where the sound field is assumed to consist of a small number of plane waves arriving from distinct directions~\cite{wabnitz2012frequency, epain2013super, routray2020sparse, wabnitz2011upscaling,kentgens2021upscaling}. Under this assumption, the \ac{hoa} coefficients can be inferred by solving a sparse recovery problem~\cite{eldar2012compressed}. \RI{Alternatively, parametric methods bypass explicit spatial sparsity assumptions by directly estimating the sound field's parameters and resynthesizing the sound field to a given order\cite{politis2015sector,politis2018compass}.} While such techniques can yield accurate reconstructions under ideal free-field conditions with limited noise and few active sources, their performance degrades significantly in challenging acoustic environments when the underlying assumptions do not hold.
% In particular, reverberation and diffuse sound components violate the directional sparsity assumption~\cite{zhang2017surround}, making the inverse problem substantially more challenging and often leading to degraded reconstructions.

%Several methods have been proposed in the literature for \ac{au}. Early approaches are predominantly model-based and rely on strict physical assumptions. A representative category applies \ac{cs} techniques for \ac{pwd} under the assumption that the sound field is sparse in the source domain \cite{wabnitz2012frequency, routray2020sparse, wabnitz2011upscaling}. While \ac{cs} techniques enable upscaling under ideal free-field and low-noise settings, their performance deteriorates significantly when the sound field becomes complex and the sparsity assumption no longer holds.

Motivated by the success of deep learning in natural language processing and computer vision, recent studies have explored data-driven paradigms for spatial audio enhancement, spanning both discriminative and generative frameworks. Representative discriminative approaches include the study in~\cite{zang2024ambisonizer}, which proposed a transformer-based architecture to synthesize Ambisonic signals from stereo inputs, and the usage of \aclp{pinn}  suggested in~\cite{miotello2024physics} to upscale spherical microphone array recordings by enforcing wave-equation consistency. 
In parallel, the emergence of generative modeling has introduced new capabilities for spatial synthesis. Early efforts in this direction, such as the work by Xia \emph{et al.}~\cite{xia2023upmix}, utilized \acp{gan} to upscale \ac{foa} \acp{rir}, aiming to recover high-resolution spatial characteristics from room impulse responses. Building on this, Picard \emph{et al.}~\cite{picard2026sirup} employed diffusion models to upscale the steering vectors of \ac{foa} signals, while Kushwaha \emph{et al.}~\cite{kushwaha2025diff} demonstrated the generation of \ac{foa} content directly from textual descriptions. Furthermore, Liang \emph{et al.}~\cite{liang2026immersiveflow} utilized \ac{fm} to synthesize 7.1.4 surround sound layouts from stereo sources. 

Specifically for the task of \ac{au},  mostly discriminative approaches have been introduced. For instance, Gao \emph{et al.} \cite{gao2022sparse} proposed a multi-scale \ac{cnn} operating in the frequency domain with sparse encoding, while Routray \emph{et al.} \cite{routray2019deep} presented a multi-stage \ac{dnn} architecture to upsample signals order-by-order. Recently, Chatzimoustafa \emph{et al.} \cite{chatzimoustafa2025higher} refined this hierarchical approach by replacing linear layers with \acp{gru} to better capture temporal dependencies. However, these models, including more recent encoder-decoder architectures like adapted \ac{ctn} \cite{nawfalambisonics}, primarily demonstrate success in anechoic or low reverberation conditions.
Generative modeling has to date only been considered for \ac{au} in the  work from 2021 by Zhang \emph{et al.}~\cite{zhang2021optimization}, which explored the use of early generative architectures based on \acp{gan} for this task, focusing on  single-tone signals from specific directions. %As   providing only a baseline comparison against the least-squares method. Our preliminary work~\cite{milstein2025diffau} introduced a diffusion-based framework that demonstrated clear superiority over both discriminative \ac{ctn} architectures~\cite{nawfalambisonics} and traditional \ac{cs} approaches. While effective in anechoic scenarios, that framework did not address the challenges of complex reverberant environments, where stochastic late reflections significantly degrade spatial reconstruction.

Despite these advances in \ac{au}, a major challenge remains the reconstruction of spatial details in reverberant environments. In such scenarios, late reflections generate spatially complex, diffuse-like sound fields that deviate significantly from the directional sparsity assumptions underlying many model-based techniques~\cite{zhang2017surround}. At the same time, discriminative learning methods that rely on deterministic mappings~\cite{shlezinger2022discriminative} from low- to high-order representations often struggle to capture the complex and stochastic spatial structure introduced by reverberation. As a result, existing methods frequently produce spatially blurred reconstructions in which distinct directional cues are not well resolved, leading to a loss of high-frequency spatial detail and reduced perceptual immersion.

 % \textcolor{red}{A short literature review recovered a few works that use generative learning tools for tasks related to \ac{au}. These include GAN based methods \cite{zhang2021optimization,xia2023upmix}, diffusion models~\cite{picard2026sirup,kushwaha2025diff}, and even flow marching~\cite{liang2026immersiveflow}. There are also a few other relevant discriminative models such as \cite{zang2024ambisonizer, miotello2024physics}. We need to include them in the literature review on related deep learning methods, while concluding with what is missing and motivating our work }

%The most significant hurdle for these discriminative models remains the reverberant scenario. In such environments, late reflections create a diffuse sound field that violates the directional sparsity these models rely on. This typically results in a spatial blurring effect, where the model fails to resolve distinct directional cues, leading to a loss of high-frequency spatial detail and immersion.

%\Amit{Propose vs extend?}
To address the limitations of existing approaches, we introduce a generative framework for \ac{au} based on state-of-the-art {\em continuous-time generative modeling}. Our approach casts  \ac{au} as sampling from the conditional distribution of missing high-order spatial components given the observed low-order Ambisonic signals. Unlike conventional approaches that rely on deterministic mappings from low- to high-order representations, our formulation treats the reconstruction of high-order harmonics as a generative inference problem, allowing the model to capture the stochastic spatial structures that arise in reverberant sound fields. To extend \ac{au} into a generative paradigm for challenging acoustic conditions, we investigate {continuous-time generative formulations} that provide improved training stability and sampling efficiency. The proposed framework combines flexible architectural design with physically motivated reconstruction constraints to generate spatially consistent \ac{hoa} representations while preserving the reliability of the measured low-order components.

Our main contributions are summarized as follows:
\begin{itemize}
    \item \textbf{Generative formulation for \ac{au}:}  
    We introduce a generative perspective for \ac{au}, in which the missing \ac{hoa} components are recovered by modeling their conditional distribution given the observed low-order signals. This formulation enables the reconstruction process to account for the inherent ambiguity and stochasticity of spatial sound fields in reverberant environments.

    \item \textbf{Continuous-time generative modeling for spatial audio:}  
    We investigate two dominant continuous-time generative paradigms for the \ac{au} task, namely \ac{sgm}~\cite{song2020score} and \ac{fm}~\cite{lipman2022flow}. While \ac{sgm} reconstructs the target distribution through reverse diffusion driven by score estimation, \ac{fm} learns a deterministic vector field that transports samples from a noise prior to the target \ac{hoa} distribution, offering improved stability and sampling efficiency.

    % \item \textbf{Flexible  upscaling architecture:}  
    % We develop a flexible generative architecture that can operate either as a single-stage model or as a cascaded multi-stage system that progressively increases the Ambisonic order. We analyze the trade-offs between these configurations and study their impact on spatial coherence and error propagation in reverberant environments. \NirCmt{we may need to reword this contribtuion}

    \item \textbf{Targeted reconstruction of high-order spatial components:}  
    We introduce a reconstruction strategy that preserves the measured low-order Ambisonic components while restricting the generative model to synthesize only the missing high-order channels. This design leverages the physical reliability of the observed data while focusing the generative capacity on resolving the underdetermined spatial details.

    \item \textbf{Extensive evaluation:}  
    We conduct a comprehensive experimental study, including objective metrics and subjective listening tests across varied acoustic scenarios, demonstrating that the proposed generative framework improves the reconstruction of spatial details compared to existing discriminative baselines, particularly in challenging reverberant scenarios.
\end{itemize}

The rest of this paper is organized as follows. Section~\ref{sec:Preliminaries} introduces the signal model for Ambisonics and provides the mathematical foundation for continuous-time generative models. Section~\ref{sec:State} details our proposed generative framework, describing the flexible hierarchical architecture, the targeted reconstruction strategy, and the signal preprocessing pipeline. The empirical validity of the approach is evaluated in Section~\ref{sec:emp_eval}, where we present a comprehensive numerical study across varied reverberant conditions followed by the results of our subjective listening tests in Sections~\ref{ssec:listening_exp_I} \RI{and \ref{ssec:listening_exp_II}}. Finally, Section~\ref{sec:conclusions} provides concluding remarks.

% Notations....

Throughout this paper, we denote multi-dimensional quantities using boldface notation, with their dimensions specified upon introduction. %The $(i,j)$th entry of  $\myMat{X}$ is denoted by $[\myMat{X}]_{i,j}$. 
Calligraphic letters denote sets, e.g., $\mySet{X}$,  with $\mathbb{R}$ and $\mathbb{C}$ being the sets of real and complex numbers,  while  $\|\cdot\|$, $\|\cdot\|_F$,  $(\cdot)^T$, $(\cdot)^H$, and $(\cdot)^{\dagger}$   are the $\ell_2$ norm, Frobenius norm, transpose,  conjugate transpose, and Moore-Penrose pseudo-inverse, respectively.   For tensor indexing, $[\myMat{X}]_{i_1,i_2,\dots,i_n}$ denotes a scalar entry of an $n$-dimensional tensor, while 
$[\myMat{X}]_{:,\dots,:,i_k,:,\dots,:}$ denote sub-tensors obtained by fixing the $k$th index.

%%%%%%%%%%%%%%%%%%%%
%%%	Preliminaries %%%
%%%%%%%%%%%%%%%%%%%%
%
% \vspace{-0.2cm}
\section{Preliminaries and Problem Formulation}\label{sec:Preliminaries}
% \vspace{-0.15cm}
This section provides the  background for \ac{au}. We begin by formulating the signal model and the fundamental properties of Ambisonics signals in Subsection~\ref{subsec:Signal_Model}, followed by the formulation of the \ac{au} task as an inverse problem in Subsection~\ref{subsec:problem_formulation}. Then, to motivate our methodology, we review the preliminaries of continuous-time generative models in Subsection~\ref{ssec:gen_Preliminaries}.

\subsection{Signal Model and Ambisonics}\label{subsec:Signal_Model} 
Consider an array comprising $M$ omni-directional microphones positioned at coordinates $(r_i, \theta_i, \phi_i)$ for $1 \leq i \leq M$. The surrounding sound field is assumed to be composed of $Q$ plane waves arriving from directions $(\theta_q, \phi_q)$, where $q \in \{1, \ldots, Q\}$. The signal captured by the array at wave number $k$ is expressed as
\begin{equation}\label{eq:mic_equation}
    \myVec{p}(k) = \myMat{V}(k) \myVec{\xi}(k) + \myVec{n}(k),
\end{equation}
where $\myVec{p}(k) = [p_1(k), \ldots, p_M(k)]^T$ is the $M \times 1$ vector of measured sound pressures, and $\myMat{V}(k)$ is the $M \times Q$ array steering matrix whose $(i,q)$-th element represents the frequency response of the $i$-th microphone to a plane wave from direction $q$. The vector $\myVec{\xi}(k) = [\xi_1(k), \ldots, \xi_Q(k)]^T$ represents the $Q$ source signals at the origin, and $\myVec{n}(k)$ denotes additive i.i.d. microphone noise.

The Ambisonic representation of order $N$ for the $Q$ plane waves $\myVec{\xi}(k)$ is given by~\cite{rafaely2015fundamentals}
\begin{equation}\label{eq:ambi_signal_ideal}
    \myVec{a}_{N}(k) = \myMat{Y}_Q^H \myVec{\xi}(k),
\end{equation}
where $\myVec{a}_{N}(k)$ is a vector of size $(N+1)^2$ containing the Ambisonic components. The matrix $\myMat{Y}_Q \in \mathbb{C}^{Q \times (N+1)^2}$ is comprised of the spherical harmonic functions sampled at the source directions $(\theta_q, \phi_q)$. To prevent spatial aliasing, the condition $k \cdot r \ll N$ must be satisfied, 
while $(N + 1)^2 \leq M$ is required to facilitate Ambisonics encoding from microphone signals.

In practice, the plane wave signals $\myVec{\xi}(k)$ are not directly accessible, and the Ambisonic components are acquired from the measured $\myVec{p}(k)$. In the case of a spherical microphone array, the steering matrix $\myMat{V}(k)$ in \eqref{eq:mic_equation} is decomposed as:
\begin{equation}
    \myMat{V}(k) = \myMat{Y}_M \myMat{B}(k) \myMat{Y}_Q^H,
\end{equation}
where $\myMat{Y}_M \in \mathbb{C}^{M \times (N+1)^2}$ is the spherical harmonic matrix evaluated at the $M$ microphone positions, and $\myMat{B}(k)$ is a diagonal matrix of mode strength functions accounting for the array's radial characteristics. The  Ambisonic signal of \eqref{eq:ambi_signal_ideal} is estimated by inverting the physical model, i.e., via
\begin{equation}\label{eq:ambi_estimate}
    \hat{\myVec{a}}_{N}(k) = (\myMat{Y}_M \myMat{B}(k))^{\dagger} \myVec{p}(k).
\end{equation}
 While the representation in \eqref{eq:ambi_estimate}, termed {\em encoding}, is specific for spherical arrays, encoding techniques for general microphone arrays have been extensively studied, ranging from parametric spatial models to generalized neural encoders \cite{gayer2024ambisonics, mccormack2022parametric, heikkinen2025gen}.

%Ultimately, the order $N$ dictates the spatial resolution of the sound field representation. While increasing $N$ improves localization accuracy and reconstruction fidelity, it is physically constrained by the number of microphones $M$ and the spatial aliasing limit .% Furthermore, the numerical stability of the inversion in \eqref{eq:ambi_estimate} often degrades at low frequencies due to the ill-conditioning of $\myMat{B}(k)$, necessitating the generative upscaling approach proposed in this work.

\subsection{Problem Formulation}\label{subsec:problem_formulation} 
The Ambisonic order ($N$ in \eqref{eq:ambi_signal_ideal}) dictates the spatial resolution of the sound field representation. While increasing $N$ improves localization accuracy and reconstruction fidelity, it is physically constrained by the number of microphones $M$ and the spatial aliasing limit. 
\ac{au} is defined as the task of reconstructing a \ac{hoa} representation from a restricted low-order observation. Specifically, \ac{au} aims to map a low-order signal of order $N$ into its high-order counterpart of order $N'$, where $N' > N$. This task can be mathematically framed as a linear inverse problem in the time-domain of the form
\begin{equation}\label{eq:inverse_prob_formulation_time}
    \myVec{a}_{N}(\tau) = \myMat{F} \myVec{a}_{N'}(\tau),
\end{equation}
where $\myVec{a}_{N}(\tau) \in \mathbb{R}^{(N+1)^2}$ represents the captured time-domain low-order components and $\myVec{a}_{N'}(\tau) \in \mathbb{R}^{(N'+1)^2}$ denotes the desired time-domain high-order components. Both Ambisonics signals are the inverse Fourier transform of the frequency-domain components defined in \eqref{eq:ambi_signal_ideal} over all wavenumbers $k$. The matrix $\myMat{F} \in \mathbb{R}^{(N+1)^2 \times (N'+1)^2}$ is a truncation or selection matrix, typically structured as $\myMat{F} = [\myMat{I}_{(N+1)^2}, \myMat{0}]$, which effectively discards the high-order spatial components.

Our aim in \ac{au} is to recover the high-order components from their low-order measurements observed  via \eqref{eq:inverse_prob_formulation_time} over a time horizon $\tau_{\max}$, namely, for  $\tau \in \{0,\ldots, \tau_{\max}\}$. 
For design purposes, we assume access to a dataset comprised of multiple \ac{hoa} recordings. The dataset is written as
% \begin{equation}
%     \label{eqn:Dataset}
%     \mySet{D} = \left\{\left\{a_{N'}^{(i)}(\tau)\right\}_{\tau=0}^{\tau_{max}} \right\}_{i=1}^{|\mySet{D}|}.
% \end{equation}
\begin{equation}
    \label{eqn:Dataset}
    \mySet{D} = \left\{\myMat{A}^{(i)}_{N'}\ \right\}_{i=1}^{|\mySet{D}|}.
\end{equation}
where $\myMat{A}^{(i)}_{N'} \in \mathbb{R}^{(N^{'}+1)^2\times\tau_{\max}}$ is the stacking of the time-domain high-order components $\big\{\myVec{a}_{N'}^{(i)}(\tau)\big\}_{\tau=0}^{\tau_{\max}}$.

Because $\myMat{F}$ is a wide matrix, the system in \eqref{eq:inverse_prob_formulation_time} is  underdetermined, meaning there exists an infinite set of high-order signals that satisfy the low-order observations. Traditionally, this is addressed using the least-norm solution via the Moore-Penrose pseudo-inverse\cite{barata2012moore}. However, as noted in \cite{epain2009application}, such deterministic solutions tend to distribute energy uniformly across the spatial domain, often leading to "spatial blurring," where the directional sharpness of the original plane waves is lost, ultimately degrading the localization cues in the rendered  output.
To recover the high-frequency spatial details lost during truncation, it is essential to incorporate prior knowledge regarding the structure of typical sound fields.

\subsection{Preliminaries of Continuous-Time Generative Models }\label{ssec:gen_Preliminaries}

Continuous-time generative models are a dominant paradigm in generative learning~\cite{liu2025flowing}.
The general framework of continuous-time generative models aims to learn a smooth transformation between a tractable prior distribution, $p_0$ (typically a simple Gaussian), and a complex target data distribution, $p_1$, with both distributions defined over some vector space $\mathbb{R}^d$. This transition is conceptualized as a continuous path in probability space with time variable $t \in [0, 1]$, and specifically as a stochastic process $\{\myVec{x}_t\}_{t\in [0,1]}$ formulated via its differential $d \myVec{x}_t$ such that $\myVec{x}_0 \sim p_0$ and $\myVec{x}_1 \sim p_1$. We next elaborate on two representative forms of such generative models: {\em \ac{fm}}~\cite{lipman2022flow} and {\em \acp{sgm}}~\cite{song2019generative} .

% \textcolor{red}{the formulation is not unified and not all that clear. Can you try following the framework as in \url{https://icml.cc/virtual/2025/40011}? I started watching this tutorial and it is really good. Maybe we should actually start with FM and then go into SGM}

\subsubsection{Flow Matching}
In flow-based modeling, the transformation between distributions is defined by a transport map governed by an \ac{ode} of the form:
\begin{equation}\label{eq:fm_ode}
    d \myVec{x}_t= u_t(\myVec{x}_t)dt, \quad t \in [0,1],
\end{equation}
where ${u}_t(\cdot)$ is a time-dependent \ac{vf} that ensures the marginals of $\myVec{x}_t$ match the data distributions at $t=0$ and $t=1$. Since infinitely many paths can satisfy these boundary conditions, a common approach is to define a linear probability path
\begin{equation}\label{eq:fm_xt_interpolation}
    \myVec{x}_t = (1 - t)\myVec{x}_0 + t\myVec{x}_1,
\end{equation}
such that the conditional velocity field is expressed as
\begin{equation}\label{eq:conditional_vf}
    u_t(\myVec{x}_t | \myVec{x}_0, \myVec{x}_1) = \frac{\myVec{x}_1 - \myVec{x}_t}{1-t}.
\end{equation}
% Due to the singularity at $t=1$,
% By differentiating \eqref{eq:fm_xt_interpolation} with respect to $t$, we obtain the constant velocity $\frac{d \myVec{x}_t}{dt} = \myVec{x}_1 - \myVec{x}_0$. Using \eqref{eq:fm_xt_interpolation} to solve for $\myVec{x}_1$ and substituting it back, the conditional velocity field is expressed as:
% \begin{equation}\label{eq:conditional_vf}
%     u_t(\myVec{x}_t | \myVec{x}_0, \myVec{x}_1) = \frac{\myVec{x}_1 - \myVec{x}_t}{1-t}
% \end{equation}

The learning objective of \ac{fm} is to train a \ac{dnn}  with trainable parameters $\myVec{\theta}$, whose mapping $\myVec{v}(\myVec{x}_t, t; \myVec{\theta})$ approximates this vector field. To avoid numerical instability caused by the singularity at $t=1$ in \eqref{eq:conditional_vf}, the model is typically trained to regress the constant displacement $(\myVec{x}_1 - \myVec{x}_0)$, which is derived by differentiating \eqref{eq:fm_xt_interpolation} with respect to $t$, via the \ac{cfm} objective. This is shown in \cite{lipman2022flow} to be equivalent in expectation to the \ac{fm} objective.

Once the model $\myVec{v}(\mathbf{x}_t, t; \boldsymbol{\theta})$ is trained, samples are generated by solving the \ac{ode} in \eqref{eq:fm_ode} starting from a noise distribution $\mathbf{x}_0 \sim p_0$. Using the Euler method with $J$ discrete time steps and a step size $\Delta t = 1/J$, the trajectory is computed as in Algorithm \ref{alg:fm_sampling}.

\begin{algorithm}
\caption{\ac{fm} Sampling  via Euler Method}
\label{alg:fm_sampling} 

\SetKwInOut{Input}{Input}
\Input{Number of steps $J$ \\Prior distribution $p_0$\\ Trained model parameters $\myVec{\theta}$}

Sample $\myVec{x}_0 \sim p_0$;\\
Set step size $\Delta t \leftarrow 1/J$\;

\For{$j = 0$ \KwTo $J-1$}{
    Set $t_j \leftarrow j \Delta t$\;
    $\myVec{u} \leftarrow \myVec{v}(\myVec{x}_{t_j}, t_j; \myVec{\theta})$\;
    $\myVec{x}_{t_{j+1}} \leftarrow \myVec{x}_{t_j} + \myVec{u} \Delta t$\;
}

\KwRet{$\myVec{x}_{1}$}
\end{algorithm}

% The method  allows for the construction of "straight" probability paths using \ac{ot} displacement. By learning a vector field that defines a constant-velocity mapping between $p_0$ and $p_1$, the \ac{ode} can be solved efficiently with fewer function evaluations, facilitating faster and more stable high-fidelity reconstruction.

\subsubsection{Score-based Generative Modeling}
While \ac{fm} relies on deterministic trajectories, \ac{sgm} generalizes the continuous-time framework by framing the generative process as the solution to a reverse-time \ac{sde}. In this setting, the transformation is not purely a velocity-driven transport map, but is augmented with a stochastic term based on Langevin dynamics. The reverse-time \ac{sde} is expressed as:
\begin{equation}\label{eq:diff_sde}
    d \myVec{x}_t = \left[ {u}_t(\myVec{x}_t, t) - g_t^2 \nabla_{\myVec{x}} \log p_t(\myVec{x}_t) \right] dt + g_t d\myVec{w}_t,
\end{equation}
where $g_t$ is the diffusion coefficient, $d\myVec{w}_t$ is a reverse-time Wiener process, and $\nabla_{\myVec{x}} \log p_t(\myVec{x}_t)$ is the score function of the marginal density at time $t$. 

The learning objective is to train a \ac{dnn} $\myVec{s}(\myVec{x}_t, t; \myVec{\theta})$ to approximate the score function for all $t \in [0,1]$, while treating $u_t(\cdot)$ as a predetermined function which depends on a specific diffusion family.
This learning framework provides a unified perspective that encompasses several prominent model families of diffusion generative models, including \ac{ve}, \acl{vp}, \acl{sub-vp}, and \acl{ou} formulations, by appropriately configuring  $u_t(\cdot)$ and $g_t$ \cite{song2020score_sde, richter2023speech}. Once trained, sampling is performed using Predictor-Corrector (P-C) methods \cite[Appendix G]{song2020score_sde}, which combine numerical \ac{sde} solvers with Langevin-type MCMC updates as outlined in Algorithm \ref{alg:sgm_sampling}. During the corrector step, the Langevin step size $\delta_j$ is adaptively determined at each iteration to maintain a constant signal-to-noise ratio.

\begin{algorithm}
\caption{\ac{sgm} Predictor-Corrector Sampling}
\label{alg:sgm_sampling} 

\SetKwInOut{Input}{Input}
\Input{Number of predictor steps $J$\\
Prior distribution $p_0$\\
Trained model parameters $\myVec{\theta}$}

Sample $\myVec{x}_0 \sim p_0$\; %(e.g., $\myVec{x}_0 \sim \mathcal{N}(\myVec{0}, \myVec{I})$)\;
Set step size $\Delta t \leftarrow 1/J$\;

\For{$j = 0$ \KwTo $J-1$}{
    Set $t_j \leftarrow j \Delta t$\;
    
    \tcp{P Step: Euler-Maruyama SDE Solver}
    $\myVec{z}_p \sim \mathcal{N}(\myVec{0}, \myVec{I})$\;
    $\myVec{f} \leftarrow \big( u_{t_j}(\myVec{x}_{t_j}, t_j) - g_{t_j}^2 \myVec{s}(\myVec{x}_{t_j}, t_j; \myVec{\theta}) \big)$\;
    $\myVec{x} \leftarrow \myVec{x}_{t_j} + \myVec{f} \Delta t + g_{t_j} \sqrt{\Delta t} \myVec{z}_p$\;
   %\newline
    \tcp{C Step: Langevin MCMC}

    $\myVec{z}_{c} \sim \mathcal{N}(\myVec{0}, \myVec{I})$\;
    $\myVec{x} \leftarrow \myVec{x} + \delta_j \myVec{s}(\myVec{x}, t_{j+1}; \myVec{\theta}) + \sqrt{2\delta_j} \myVec{z}_{c}$\;
    
    % \textcolor{red}{what is $\epsilon_n$?}

    Set $\myVec{x}_{t_{j+1}} \leftarrow \myVec{x}$\;
}
\KwRet{$\myVec{x}_{1}$}
\end{algorithm}

\section{Generative Ambisonic Upscaling}
\label{sec:State}
% \vspace{-0.1cm}
% % Background 
\begin{figure*}
    \centering
    \includegraphics[width=\linewidth]{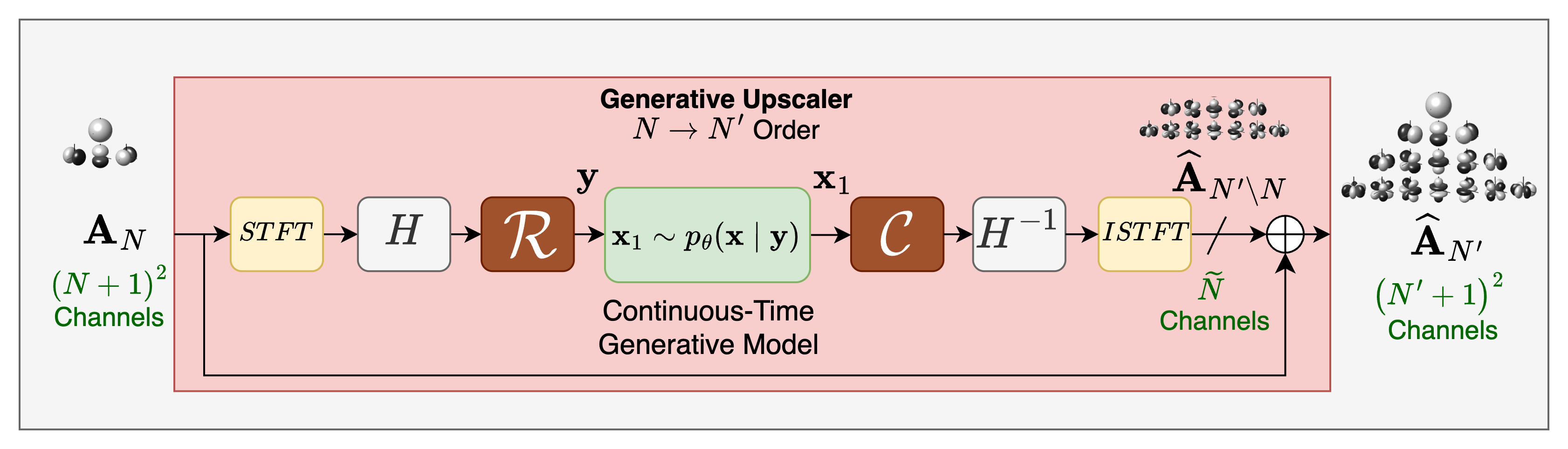}
    \caption{Schematic illustration of the overall architecture of the proposed generative Ambisonics upscaler. }
    \vspace{0.3cm}
    \label{fig:FMAU}
\end{figure*}

In this section, we present our  generative framework for \ac{au}. We begin in Subsection~\ref{ssec:ode_formulation} by formulating  the continuous-time generative models discussed in the previous section for spatial audio. Subsection~\ref{ssec:proposed_method} details our targeted reconstruction strategy, which focuses on completing missing high-order harmonics while preserving the physical integrity of the observation. The training pipeline is described in Subsection~\ref{subsec:Training}, and is followed by a discussion in Subsection~\ref{ssec:discussion}.

\subsection{Differential Equation Formulation for Ambisonics Upscaling}\label{ssec:ode_formulation}
While the framework in Subsection~\ref{ssec:gen_Preliminaries} considers generic  continuous-time generative modeling, we adapt it for the \ac{au} task. 
In our formulation, the sampled variable $\myVec{x}_t$ represents the \ac{hoa} components. 
We define a probability path that interpolates between a Gaussian noise prior $\myVec{x}_0 \sim p_0$ and the target distribution $\myVec{x}_1 \sim p_1$, which is the \ac{hoa} distribution conditioned on the observed low-order components $\myVec{y} = \TDAmbMat{N}$. 
By governing the generative process via the conditional probability path $p_t(\myVec{x} | \myVec{y})\triangleq p(\myVec{x}_t = \myVec{x} | \myVec{y})$, the model reconstructs high-order spherical harmonics directly from low-order observations, effectively sampling from $p_1(\myVec{x} | \myVec{y})$. Consequently, the reconstruction of the missing Ambisonic orders based on the inverse problem \eqref{eq:inverse_prob_formulation_time} is constrained to physically plausible solutions through posterior sampling.
This formulation specializes both the  \ac{fm} and \ac{sgm}-oriented formulations for the differential equations governing the generative process, as detailed next:

\subsubsection{\ac{fm} \ac{ode}}
In this framework, the  \ac{vf} component in the \ac{ode} \eqref{eq:fm_ode} is conditioned on the observations. Accordingly, the underlying \ac{ode} is given by
\begin{equation}\label{eq:Cond_fm_ode}
    d \myVec{x}_t= u_t(\myVec{x}_t| \myVec{y})dt, \quad t \in [0,1].
\end{equation}
%we train an \textit{observation-dependent conditional \ac{vf}} that estimates $u_t(\myVec{x}_t \mid \myVec{x}_0,\myVec{x}_1, \myVec{y})$. 
Following the linear probability path defined in \eqref{eq:fm_xt_interpolation}, the \ac{vf} should regress the constant-velocity field required to map a sample from the Gaussian prior to a high-order completion $\myVec{x}_1$, while being consistent with the low-order components $\myVec{y}$.

\subsubsection{\ac{sgm} SDE}
This setting requires a \textit{conditional diffusion model} which must estimate the conditioned score function $\nabla_{\myVec{x}} \log p_{t}(\myVec{x}|\myVec{y})$. The \ac{sde} we use is the \ac{ve} \ac{sde} introduced in\cite{song2020score_sde}. In this formulation, the \ac{vf} is fixed to $u_t(\myVec{x}_t,t)= 0$, while the diffusion term $g_t$  is set to 
\begin{subequations}
\label{eq:diffau_diff_coeff}
\begin{align}
    g_t &= \sigma_t \sqrt{2 \log \left( \frac{\sigma_{\max}}{\sigma_{\min}} \right)},
\end{align}
where $\sigma_t$ is a geometric scheduler defined as:
\begin{equation}\label{eq:sgm_noise_scheduler}
    \sigma_t = \sigma_{\min} \left( \frac{\sigma_{\max}}{\sigma_{\min}} \right)^t.
\end{equation}
\end{subequations}
Here, $\sigma_{\min}$ and $\sigma_{\max}$ are predefined noise level limits.

\subsection{Proposed Generative Framework}\label{ssec:proposed_method}

\subsubsection{High Level Architecture}
The overall architecture of generative \ac{au} is illustrated in Fig.~\ref{fig:FMAU}. While existing work on generative models for inverse problems typically focuses on spatial dimensions (in computer vision \cite{lugmayr2022repaint,chung2022diffusion}) or temporal and spectral dimensions (in audio \cite{lemercier2025diffusion,daras2024survey}), in \ac{au} the missing dimensions to be generated correspond to the  restoration of the Ambisonics channel dimension.
Specifically, we seek a \textit{targeted reconstruction} of the channel dimension, i.e., rather than generating the full $(N' + 1)^2$-channel signal from scratch, we allocate the entire generative capacity to the $\Tilde{N}\triangleq (N'+1)^2-(N+1)^2$ missing channels for orders $n \in \{N+1, \dots, N'\}$. This formulation treats the \ac{au} task as a constrained inverse problem where the low-order components are fixed observations.

To solve this, we let $\hat{\myMat{A}}_{N'\setminus N} \in \mathbb{R}^{\Tilde{N}\times\tau_{\max}}$ denote the high-order components reconstructed by the model. The final estimate $\hat{\myMat{A}}_{N'} \in \mathbb{R}^{(N'+1)^2 \times \tau_{\max}}$ is then assembled by concatenating the physically measured low-order input with these generated channels as
\begin{equation}
    \TDAmbMatest{N'} = \begin{bmatrix} \TDAmbMat{N} \\ \TDAmbMatest{N'\setminus N} \end{bmatrix}.
\end{equation}
This assembly strategy ensures that the low-order spatial information (where the physical measurement is most reliable) is fully preserved. Consequently, the generative model, which is guided by the measured low-order components $\myVec{y}$ and follows the differential equations detailed in Subsection~\ref{ssec:ode_formulation}, only needs to resolve the underdetermined portion of the inverse problem: the high-order spherical harmonics.

\subsubsection{Signal Representation}\label{ssec:signal_representation}

The input is $\TDAmbMat{N}$, a time-domain real Ambisonic signal of order $N$. We first transform the signal into the \ac{tf} domain via the \ac{stft}, and denote the resulting representation as $\FDAmb{N}$.

To address the heavy-tailed distribution of speech amplitudes and ensure consistent scaling for the \ac{dnn} implementing the generative mapping, we apply a nonlinear amplitude transformation $\mathcal{H}(x)$~\cite{richter2023speech,dong2025edsep}, given by
\begin{align}
\mathcal{H}(x) = \beta|x|^\alpha e^{j \arg(x)}, \quad
\mathcal{H}^{-1}(x) = \frac{|x|^{1/\alpha}}{\beta} e^{j \arg(x)},\label{eq:bwd_transformation}
\end{align}
where $\alpha$ and $\beta$ are compression and scaling hyperparameters, respectively. Since the \ac{dnn} operates in the real domain, we define a transformation $\mySet{R}$ that concatenates the real and imaginary components along the channel dimension. Following the generative process, an inverse transformation $\mySet{C}$ collapses these channels back into complex values, after which $\mathcal{H}^{-1}$ and the \ac{istft} are applied to recover the time-domain \ac{hoa} components. The complete \ac{au} procedure is summarized in Algorithm~\ref{alg:gen_upscaler}. While the underlying generative logic remains consistent with the frameworks described in Subsection~\ref{ssec:gen_Preliminaries}, the sampling process is now explicitly conditioned on the available low-order components $\myVec{y}$.

\begin{algorithm}
\caption{Generative Upscaling}
\label{alg:gen_upscaler} 

\SetKwInOut{Input}{Input}
\Input{
    $\TDAmbMat{N} \in \mathbb{C}^{(N+1)^2\times \Tsig}$ \\
        Number of steps $J$ \\
    Generative model parameters $\myVec{\theta}$\\
}
\BlankLine
$\FDAmb{N} \leftarrow \text{STFT}(\TDAmbMat{N})$ \;
$\myVec{y} \leftarrow \mySet{R}\{\mySet{H}(\FDAmb{N})\}$ \;

\BlankLine
% Initialize $2\tilde{N}$ channels $\myVec{x}_0 \sim p_0$ \;
$\myVec{x}_1 \leftarrow \text{Sampler}(\myVec{y},J,p_0,\myVec{\theta})$ \hfill (via Algorithm \ref{alg:fm_sampling} or \ref{alg:sgm_sampling}) \;
\label{stp:Sample}

\BlankLine
$\FDAmbest{N'\setminus N} \leftarrow \mySet{H}^{-1}(\mySet{C}(\myVec{x}_1))$; \\
$\TDAmbMatest{N' \setminus N} \leftarrow \text{ISTFT}(\FDAmbest{N'\setminus N})$ \;

$\TDAmbMatest{N'} \leftarrow {\rm Concat}(\TDAmbMat{N}, \TDAmbMatest{N' \setminus N})$ \;

\BlankLine
\KwRet{$\TDAmbMatest{N'} \in \mathbb{C}^{(N'+1)^2\times \Tsig}$}
\end{algorithm}

% \textcolor{red}{Where did you defined $\mySet{M}$? Also, the following subsection is packed with undefined notations and is not consistent with how things are formulated in the previous subsections. Please go over Sections 2-3, and then revise the training subsection to match the formulation without introducing new unnecessary variables.}

\subsection{Training}
\label{subsec:Training}
The proposed training objective aims to recover the underlying conditional probability path $\{p_t\}_{t \in [0,1]}$, which provides the generative trajectory for the sampler in Step~\ref{stp:Sample} of Algorithm~\ref{alg:gen_upscaler}. By approximating this path, the model ensures that the generated $\myVec{x}_1$ constitutes a faithful recovery of the \ac{hoa} components. Specifically, the learnable parameters $\myVec{\theta}$ are used to estimate the score function in \acp{sgm} or the \ac{vf} in \ac{fm}. In the following, we first describe the data preparation procedure, followed by the specific training losses for both frameworks.

\subsubsection{Data Preparation}
To this end, we sample a time-domain Ambisonic signal $\TDAmbMat{N'}$ of order $N'$ from the dataset $\mySet{D}$ in \eqref{eqn:Dataset} and partition it into its  low- and high-order components. Specifically, we define the conditioning signal $\myVec{y}$ using the first $(N+1)^2$ channels, corresponding to the low-order signal $\TDAmbMat{N}$. The remaining $\tilde{N}$ channels, representing the high-order components $\TDAmbMat{N' \setminus N}$, are designated as the target data sample $\myVec{x}_1$.

\subsubsection{Training Objective}
Due to the intractability of the continuous-time objective, these models are trained using Monte Carlo estimation. Specifically, for each $(\myVec{x}_1, \myVec{y})$ pair, we draw a time step $t \sim \mathcal{U}(0, 1)$ and a noise sample $\myVec{x}_0 \sim p_0$. We denote the collection of $|\mySet{D}|$ sampled time steps and noise samples as $\myVec{t}$ and $\myMat{X}_0$, respectively. Based on these samples, the specific training objective varies by framework. \RI{The proposed training objectives for both \ac{fm} and \ac{sgm} operate on multiple channels, allowing the network to implicitly learn the underlying physical inter-channel dependencies. While one can in principle introduce additional regularization and explicitly encourage inter-channel correlation, we utilize the loss functions detailed in the sequel based on our empirical exploration and their established usefulness in training continuous-time generative models.}

\paragraph{\acl{fm}}
In this framework, $\myVec{x}_0$ which is sampled from $p_0 \equiv \mathcal{N}(\myVec{0},\myVec{I})$, serves as the source of a generative path terminating at the data sample $\myVec{x}_1$. To learn the corresponding \ac{vf}, we adopt the \textit{denoiser loss} formulation \cite{kim2024simple}.

Specifically, the interpolated state $\myVec{x}_t$ is constructed via \eqref{eq:fm_xt_interpolation}, the model then predicts the target data sample $\myVec{x}_1$ from the current state $\myVec{x}_t$ through the estimated \ac{vf}:
\begin{equation}
    \hat{\myVec{x}}_{1,t} = \myVec{x}_{t} + (1 - t) \cdot \myVec{v}(\myVec{x}_{t}, t \mid \myVec{y}; \myVec{\theta}).
\end{equation}
The resulting weighted empirical risk is given as:
\begin{equation}\label{eq:fm_loss}
    \mathcal{L}_{\mySet{D},\myVec{t},\myMat{X}_0}^{\text{CFM}}(\myVec{\theta}) = \frac{1}{|\mySet{D}| \Tilde{N}} \sum_{i=1}^{|\mySet{D}|} w(t^{(i)}) \cdot \left\| \myVec{x}_1^{(i)} - \hat{\myVec{x}}_{1,t^{(i)}}^{(i)} \right\|^2.
\end{equation} 
While a weighting of $w(t) = (1-t)^{-2}$ recovers the original velocity-regression objective of \cite{lipman2022flow}, we employ a logarithmic scheme $w(t) = \log(1+t)$, inspired by practices employed in training deep unfolded models~\cite{shlezinger2025deep}. Empirically, this was found to enhance training stability and generative quality.

\paragraph{\acl{sgm}}
% For consistency with diffusion literature, we may also refer to this noise sample as $\myVec{z}^{(i)} = \myVec{x}_0^{(i)}$. 

This approach adopts the \textit{denoising score matching} strategy \cite{vincent2011connection} to learn the score of the conditional marginal density, i.e., $\nabla p_t(\myVec{\myVec{x}\mid \myVec{y)}}$. 
The interpolated state is parameterized with the \ac{ve} perturbation kernel:
\begin{equation}\label{eq:ve_prob_path}
    \myVec{x}_{t}= \myVec{x}_1 + \sigma_{t} \myVec{x}_0,
\end{equation}
where the scheduler $\sigma_t$, as defined in \eqref{eq:sgm_noise_scheduler}, interpolates between the data distribution at $t=1$ and the prior at $t=0$, which is  $p_0 \equiv \mathcal{N}(\myVec{0}, \sigma_{\max}^2 \myMat{I}) $ \cite{shaked2025ai}. The empirical loss is defined as:
\begin{equation}\label{eq:sgm_loss}
    \mathcal{L}^{\text{SGM}}_{\mySet{D},\myVec{t},\myMat{X}_0}(\myVec{\theta})\! =\! \frac{1}{|\mySet{D}|\tilde{N}}\sum_{i=1}^{|\mySet{D}|}\left\| \myVec{s}\big(\myVec{x}_{t^{(i)}}^{(i)},t^{(i)}|\myVec{y}^{(i)}; \myVec{\theta}\big)\cdot\sigma_{t^{(i)}} \!+\! \myVec{x}_0^{(i)} \right\|^2.
\end{equation}

\subsubsection{Training Procedure}

The overall training procedure based on mini-batch \ac{sgd} is detailed in Algorithm \ref{alg:GenAU_train}. The formulation of Algorithm~\ref{alg:GenAU_train} accommodates both \ac{fm} and \ac{sgm} for generative modeling. There, we also maintain a shadow copy of the model weights, used  for inference, which is updated  via \ac{ema} with forgetting coefficient $\gamma \in [0,1]$ to improve generalization and robustness.

\begin{algorithm}
    \caption{Training Generative \ac{au} via SGD}
    \label{alg:GenAU_train}
%    \SetAlgoLined
    \SetKwInOut{Initialization}{Init}
    \Initialization{Model parameters $\myVec{\theta}$; $\myVec{\theta}_{\rm EMA} = \myVec{\theta}$\\ 
    Learning rate $\eta$; EMA weight $\gamma$}
    \SetKwInOut{Input}{Input}
    \Input{Training set  $\mySet{D} = \{\myMat{A}^{(i)}_{N'}\}_{i=1}^{|\mySet{D}|}$}  
    Partition data into high- and low-order $\mySet{D} = \{\myVec{x}_1^{(i)}, \myVec{y}^{(i)}\}$\;
    \For{${\rm epoch} = 0, 1, \ldots $}{%
        Divide  $\mathcal{D}$ into $B$ batches $\{\mathcal{D}_b\}_{b=1}^B$; \\     
        \For{$b = 1, \ldots, B$}{
           % \For{each $\TDAmbMat{N'}$}{
               % Partition to high and low order components - $(\myVec{x}_1,\myVec{y})$; \\
                Draw $|\mySet{D}_b|$ i.i.d. time steps $ \sim \mathcal{U}(0, 1)$ as $\myMat{t}$\;
                Draw $|\mySet{D}_b|$ i.i.d. samples $ \sim p_0$ as $\myMat{X}_0$\;
              %  Calculate interpolated state $\myVec{x}_t$ via \eqref{eq:fm_xt_interpolation} or \eqref{eq:ve_prob_path}; \\ 
               % Calculate $\myVec{v}(\myVec{x}_t,t \mid \myVec{y};\myVec{\theta})$ or $\myVec{s}(\myVec{x}_{t},t \mid \myVec{y}; \myVec{\theta})$;\\
           % }
            Compute loss $\mySet{L}_{\mySet{D}_b,\myVec{t},\myVec{X}_0}(\myVec{\theta})$ via \eqref{eq:fm_loss} or \eqref{eq:sgm_loss}\; 
            Set $\myVec{\theta} \leftarrow \myVec{\theta} - \eta \nabla\mySet{L}_{\mySet{D}_b,\myVec{t},\myVec{X}_0}(\myVec{\theta}) $\;
            %Backpropagate $\mySet{L}_{\mySet{D}_q,\myVec{t},\myVec{X}_0}$ and update $\myVec{\theta}$ via Adam; \\
            Set $\myVec{\theta}_{\rm EMA}  \leftarrow \gamma \myVec{\theta}  + (1-\gamma)  \myVec{\theta}_{\rm EMA} $\\
    }
  }
  \KwRet{$\myVec{\theta}_{\rm EMA} $}

\end{algorithm}

\subsection{Discussion}\label{ssec:discussion} 
The proposed methodology introduces a novel approach to \ac{au}  that is particularly well-suited to the formulation of the task in Subsection~\ref{subsec:problem_formulation}, which is inherently underdetermined. Unlike deterministic approaches that seek a single point estimate, often leading to spatial blurring, the formulation in Subsection~\ref{ssec:proposed_method} recasts \ac{au} as conditional generative inference. By modeling the conditional distribution of the missing high-order components given the observed low-order signal, the framework explicitly captures the ambiguity and stochasticity induced by reverberation. The targeted reconstruction strategy further aligns with the physical structure of Ambisonics by preserving the reliably measured low-order components while restricting the generative process to the missing channels.  As a result, the generated \ac{hoa} signals remain consistent with both the observed sound field and the underlying physical constraints, while enabling the synthesis of spatial features that reflect the true data distribution. The empirical results presented in Section~\ref{sec:emp_eval} corroborate this design, demonstrating that the proposed approach achieves accurate and robust upscaling across a wide range of acoustic conditions, including highly reverberant environments.

Beyond the specific instantiations considered in this work, the proposed formulation naturally admits several extensions. While we focus on instances of \ac{fm} and Diffusion-based models, the framework is compatible with a broad class of continuous-time generative approaches. Moreover, the current formulation assumes static acoustic scenes; extending it to account for temporal dynamics, such as moving sources or listener motion, constitutes a promising direction. Another avenue is to move beyond signal reconstruction objectives and directly optimize for downstream perceptual tasks, such as binaural rendering, thereby aligning the generative process with end-use criteria. Finally, although the results indicate that a single trained model generalizes well across related acoustic conditions, further improvements may be achieved by enabling adaptation to diverse environments, for example via online updating mechanisms or hypernetwork-based conditioning. We leave these directions for future work.

%The proposed methodology introduces a novel approach to the \ac{au} task by leveraging generative frameworks. Prior works in this domain typically rely on deterministic models (utilizing linear or non-linear mapping functions) that prioritize the minimization of point-wise reconstruction errors. However, these methods often lack access to the underlying conditional distribution of the high-order components. By formulating \ac{au} as a generative inverse problem, we address this underdetermined task through the inherent strengths of generative-based models. By modeling the conditional probability path of the missing channels relative to the \ac{foa} observations, the framework generates high-order spatial features that are statistically consistent with the measured sound field while adhering to the underlying physical properties of Ambisonic signals. In the following section, we empirically validate this framework across diverse acoustic scenarios to quantify the trade-off between this generative flexibility and perceptual accuracy.

%Mobility and moving speakers; task-based upscaling for, e.g., bineural reproduction; combination with hypernetworks or online adaptation to accomodate multiple scenarios; considering alternative generative models

% 
%----------------------------------------------------------------------------------------
%	EXPERIMENTAL STUDY
%----------------------------------------------------------------------------

% \vspace{-0.2cm}
\section{Empirical Evaluation}
\label{sec:emp_eval}

In this section, we evaluate the performance of the proposed continuous-time generative models for  \ac{au}\footnote{The source code is available at \url{https://github.com/Amitmils/GenAU}.}. The objective is to upscale a \ac{foa} signal ($N=1$) to a 3rd-order representation ($N' = 3$) denoted as \ac{hoa}3. The evaluation is structured as follows: Subsection~\ref{ssec:training_data} outlines the experimental setup and the datasets used for training and evaluation. Subsection~\ref{ssec:methodology} details the methodology, including the model configurations, the \ac{au} baselines, and the evaluation metrics. Finally, Subsection~\ref{ssec:quan_eval} reports the results of our numerical analysis.

\subsection{Experimental Setup}\label{ssec:training_data}
\subsubsection{Training Data}
\label{sssec:data}
We utilize a speech dataset comprising 100 hours of audio, where each sample is 2~s in length and sampled at 16~kHz. In total, we employed 7,600 \ac{sh-rir} derived from two distinct datasets:
\begin{itemize}
    \item \textbf{MOTUS} \cite{gotz2021dataset}: A collection of real-world recordings comprising over 800 room configurations within a $4.9\text{ m} \times 4.4\text{ m} \times 2.9\text{ m}$ space. For each configuration, four \acp{sh-rir} were recorded from fixed source directions. To ensure a vast diversity of acoustic scenes, the internal geometry was randomized by repositioning furniture and altering wall and floor materials, resulting in \ac{rt} primarily concentrated between 500~ms and 1.2~s.
    \item \textbf{HARP} \cite{saini2025harp}: A dataset of simulated \acp{sh-rir} with \ac{rt} concentrated between 200~ms and 1.5~s. These \acp{sh-rir} were generated in shoebox geometries via the image source method (reflection order 30) using the HARP codebase which utilizes the \texttt{pyroomacoustics} library \cite{Scheibler2018}. The dataset encompasses 39 rooms with volumes ranging from small ($30\text{ m}^3$) to large ($850\text{ m}^3$). While the source and receiver were maintained at a constant 1~m distance, azimuth and elevation angles were randomized. To ensure acoustic consistency, the source was placed at least 0.5~m from any wall, and the receiver position remained fixed for all simulations within a specific room configuration.
\end{itemize}
For each room configuration, we randomly selected between one and four \acp{sh-rir} and convolved them with dry audio samples from the WSJ0 corpus~\cite{garofolo1993csr}. To enhance model robustness, a random gain between $-6$~dB and $6$~dB was applied to each source. Furthermore, during training, we performed random rotations on the sound field to ensure the model achieves rotation invariance.
All signals were processed in 3rd-order Ambisonics following the N3D~\cite{zotter2019ambisonics} normalization convention. 

\subsubsection{Evaluation Data}\label{ssec:eval_data}
\begin{figure}
    \centering
    \includegraphics[width=1\linewidth]{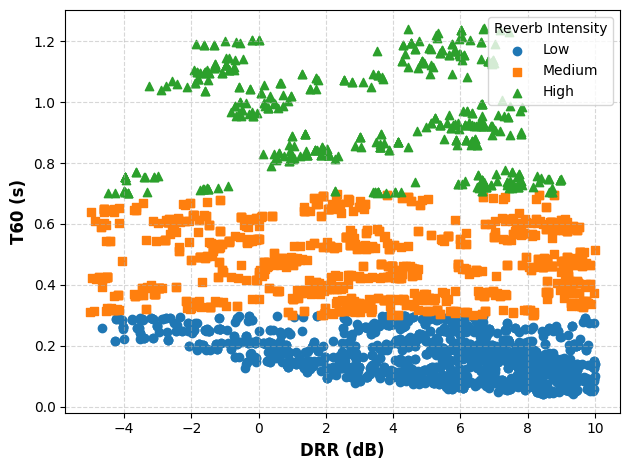}
    \caption{HARP evaluation data distribution}
    \label{fig:harp_eval_dist}
\end{figure}

We evaluate the proposed models on unseen room configurations from the HARP and MOTUS datasets, ensuring zero overlap between training and evaluation environments. 
To facilitate a comprehensive performance analysis, we simulate with HARP codebase a wide range of acoustic scenarios with \ac{rt} values from 0.1~s to 1.25~s and \ac{drr} levels from -5~dB to 10~dB. This 2-hour recording set, whose distribution is shown in Fig.~\ref{fig:harp_eval_dist}, is partitioned into subsets to analyze performance across different physical metrics. 
Specifically, the \ac{rt} subsets are categorized as \textit{Low} (\ac{rt} $< 300$~ms),  \textit{Medium} ($300 \text{ ms} \leq$ \ac{rt} $< 700$~ms), and \textit{High} (\ac{rt} $\geq 700$~ms). Similarly, the \ac{drr} subsets are divided into \textit{Low} ($\text{DRR} < 0$~dB),  \textit{Medium} ($0 \text{ dB} \leq \text{DRR} < 5$~dB), and \textit{High} ($\text{DRR} \geq 5$~dB).

Complementing this broad analysis, we also evaluate on the MOTUS dataset. This set provides a more focused benchmark for a constrained range of acoustic conditions, specifically featuring rooms with \ac{rt} between 0.5~s and 1.0~s and \ac{drr} values between -5~dB and 0~dB.
Furthermore, to evaluate robustness to the number of sources, we partitioned our datasets by the number of speakers, which ranges from 1 to 4.

As model-based \ac{au} approaches assume free-field scenarios, we also consider anechoic settings. There,  we randomly selected $1-4$ speakers for each signal, and for each selected speaker, a random \ac{doa} was assigned, and the corresponding \ac{hoa} was constructed via \eqref{eq:ambi_signal_ideal}. It is emphasized that we do not retrain the data-driven algorithms differently in anechoic settings, and use the same models trained using the reverberant dataset detailed in Subsection~\ref{sssec:data} for all scenarios. %For the \ac{fm} framework, we use the sampler 

\subsection{Methodology}\label{ssec:methodology}
\RI{We implement our generative framework utilizing the \ac{ncsn} backbone~\cite{song2020score_sde} with modifications to support multi-channel inputs and outputs specifically tailored to the \ac{au} configuration as detailed in Subsection~\ref{ssec:proposed_method}. To accommodate the high capacity required by generative frameworks to accurately model the complex data distribution of multi-channel sound fields, we employ a hidden dimension of 256, resulting in 262M parameters. While this represents a higher parameter footprint than the more compact, deterministic baselines evaluated, this capacity is essential for capturing the intricate spatial and phase relationships necessary for high-fidelity generation of complex reverberant scenarios.} This unified backbone is employed for both the \ac{sgm} (where it implements the score \ac{dnn} $\myVec{s}(\cdot)$), termed {\em Diffusion}, and \ac{fm} (where it estimates the \ac{vf} $v(\cdot)$), termed {\em Flow Matching}. For the signal representation, we employ $\alpha = 0.5$ and $\beta=0.15$ to compress and scale the signal before and after the network, respectively. \RI{We use $J=3$ step in the \ac{fm} sampler (Algorithm~\ref{alg:fm_sampling}), resulting in $3$ function evaluations, and $J=25$ steps for the \ac{sgm} sampler (Algorithm~\ref{alg:sgm_sampling}), which requires $50$ function evaluations.}
\subsubsection{Benchmarks}\label{sssec:benchmarks}
We compare our generative framework to leading data-driven and model-based \ac{au} algorithms:
\begin{itemize}
    \item {\em \ac{ctn}}: Based on the discriminative model proposed in \cite{nawfalambisonics}. We utilize a configuration with 384 encoder channels and a single repetition of 256 channels \RI{which totals in 3.5M parameters}. The model accepts \ac{foa} signals as input and targets the reconstruction of missing \ac{hoa}3 channels ($\Tilde{N}=12$), operating in a manner analogous to our proposed framework.
    \item {\em Cascaded \ac{gru}}: The model proposed in \cite{chatzimoustafa2025higher} utilizes cascaded \acp{gru} corresponding to each Ambisonic order. Each stage consists of a single \ac{gru} unit followed by a linear layer that projects the output to the required number of channels for that specific order. Following the original architecture, the hidden dimension is set to 128. We evaluate two variants: the unidirectional version as originally proposed, and a bidirectional implementation, which grants the model a non-causal advantage consistent with the other methods under test. We shall denote them as \ac{gru-u} and \ac{gru-b}, \RI{which in total have 0.2M and 0.4M parameters, respectively}.
    \item {\em PWD CS}: The model-based  framework  proposed in \cite{wabnitz2011upscaling} which casts \ac{au} as \ac{cs}. This approach minimizes the $L_{1,2}$-norm of the signals in the plane-wave domain to promote spatial sparsity. The reconstruction is subject to \ac{foa} consistency constraints, which are incorporated via Lagrange multipliers and solved using gradient descent.
    \RI{\item {\em HO-DirAC}: A parametric method proposed in~\cite{politis2015sector} that assumes a single source per \ac{tf} tile, estimating one \ac{doa} and diffuseness from \ac{foa} signals to steer \ac{hoa} covariance-matched resynthesis.}
    \RI{\item {\em COMPASS}: A parametric method proposed in~\cite{politis2018compass} that accommodates multiple simultaneous sources per \ac{tf} tile, estimating multiple \acp{doa} and diffuseness parameters to reconstruct the target \ac{hoa} components via adaptive mixing.}

\end{itemize}
 To ensure a fair comparison, all data-driven models were trained to convergence on the same dataset.

% To enhance data diversity and improve model robustness, we applied the following augmentation strategies:

% \begin{enumerate}[label=A\arabic*]
%     \item \textbf{Number of Sources:} Randomly selecting between 1 and 4 active sources within the sound field.
%     \item \textbf{Rotational Invariance:} Applying random rotations to each sound field to ensure the model remains rotation-invariant.
%     \item \textbf{Spatial Sampling:} Generating random azimuth and elevation angles for each simulated room, while maintaining a constant source-to-microphone distance of 1~m.
%     \item \textbf{Gain Variation:} Applying a random gain between $-6$~dB and $6$~dB to each source.
%     \item \textbf{Acoustic Diversity:} Varying the absorption coefficients of the wall surfaces in simulated environments to generate a wide range of reverberation times (\ac{rt}).
% \end{enumerate}

%\subsection{Evaluation Details}\label{ssec:implementation_details}

\subsubsection{Metrics} \label{sssec:metrics_def}
To assess signal reconstruction, we use the following metrics on the generated channels:
\begin{itemize}
    \item {\em \ac{stft-sdr}},  defined as 
\begin{align}
\hspace{-0.5cm}\text{STFT-SDR}(\TDAmbMatest{3\setminus1}) \!=\!
 10 \cdot \log_{10} \left( 
\frac{\|\FDAmb{3\setminus1}   \|_F^2}{ \|\FDAmb{3\setminus1} \!-\! \FDAmbest{3\setminus1} \|_F^2}
\right),
\label{eqn:STFTSDR}
\end{align} 
where $\FDAmb{3\setminus1}$ and $\FDAmbest{3\setminus1}$ represent the ground truth and estimated high-order components (2nd and 3rd order) of the Ambisonic signals in the \ac{tf} domain. A higher STFT-SDR value indicates better signal reconstruction. 
\item {\em Upscaling coherence}:   to evaluate the reconstruction quality across the frequency spectrum, we calculate the magnitude-squared coherence between the upscaled signal and the ground truth, averaged across the \ac{hoa} components.  The coherence at frequency $f$  is defined as:
\begin{equation}
\label{eqn:coherence}
\hspace{-0.5cm}
\bar{\gamma}_f^2(\TDAmbMatest{3\setminus1})\! = \!
 \frac{1}{12} \sum_{n=0}^{11} \frac{\left|  \big[\hat{\myMat{A}}^{\rm tf}_{3\setminus1}\big]_{n,:,f}^H \big[\FDAmb{3\setminus1}\big]_{n,:,f}  \right|^2}{\big\| \big[\FDAmb{3\setminus1}\big]_{n,:,f} \big\|^2 \cdot \big\| \big[\FDAmbest{3\setminus1}\big]_{n,:,f} \big\|^2 }.
\end{equation}
 This metric evaluates spectral fidelity as a function of frequency, where a value of 1 represents full reconstruction.
\end{itemize}

While \eqref{eqn:STFTSDR}-\eqref{eqn:coherence} measure  signal reconstruction fidelity, they do not fully characterize the spatial resolution of the upscaled sound field. To this end, we investigate the spatial accuracy of the upscaling in single speaker scenarios using two key spatial metrics: 
\begin{itemize}
    \item {\em \Ac{di} error}:
The \ac{di} \cite{zotter2022all} quantifies the ability to extract a directional source from a diffuse field. % Following the definition of the Directivity Factor $Q$ as the ratio of peak-to-average spatial gain, the \ac{di} 
For an order-$N$ Ambisonic signal $\TDAmbMat{N}$, the \ac{di} is
\begin{equation}
\text{DI}(\TDAmbMat{N}) = 10 \cdot \log_{10} \left( \frac{\max_{\Omega} \|\myVec{v}(\Omega)^T \TDAmbMat{N}\|^2}{\frac{1}{4\pi} \|\TDAmbMat{N}\|_F^2} \right)
\end{equation}
where $\myVec{v}(\Omega) \in \mathbb{R}^{(N+1)^2}$ is the N3D steering vector consisting of real spherical harmonics up to order $N$, for direction $\Omega = (\theta, \phi)$, with $\theta$ and $\phi$ denoting the colatitude and azimuth, respectively. %The numerator represents the maximum beam intensity, while the denominator calculates the mean intensity based on the Ambisonic coefficients.
%
%Under ideal conditions, the theoretical maximum for an $N$-order signal is $\text{DI}_{max} = 10 \cdot \log_{10}((N+1)^2)$\cite{zotter2022all}. However, in reverberant environments, the presence of a non-ideal diffuse field naturally reduces the attainable directivity of the ground-truth signal. To evaluate the upscaling accuracy relative to these physical constraints, we define the 
The \ac{di} error  is computed as 
\begin{equation}
    \Delta \text{DI}(\hat{\myMat{A}}_N) = \big|\text{DI}(\hat{\myMat{A}}_N) - \text{DI}({\myMat{A}}_N)\big|.
    \label{eqn:DIError}
\end{equation}
This metric penalizes both spatial under-estimation (blurring) and over-estimation (artificial de-reverberation), where a lower value indicates better spatial reconstruction fidelity.
\item \textit{Angular difference} ($\Delta\sigma$), defined as the great-circle distance between the \ac{doa} of the estimated and ground-truth signals. The \ac{doa} is estimated by finding the maximum intensity on a dense spherical grid:
\begin{equation}
    (\hat{\theta}, \hat{\phi}) = \underset{\Omega}{\arg\max} \, \|\myVec{v}(\Omega)^T \TDAmbMat{N}\|^2,
\end{equation}
with which the angular error is then calculated as \cite{zotter2019ambisonics}:
\begin{equation}
\Delta\sigma = \cos^{-1}\left(\sin\theta \sin\hat{\theta} + \cos\theta \cos\hat{\theta} \cos(\phi - \hat{\phi})\right),
\label{eq:angular_dist}
\end{equation}
where $(\theta,\phi)$ are the ground-truth \ac{doa} derived via an identical grid search performed on the reference \ac{hoa}3 signal. 
\end{itemize}
While $\Delta \text{DI}$ characterizes the reconstructed spatial resolution, it does not account for the directional accuracy of the soundfield. Thus, by evaluating both $\Delta \text{DI}$ and $\Delta\sigma$, we provide a holistic assessment of the model’s ability to not only sharpen the spatial response but also accurately preserve the source's  position. 

\subsection{Results}\label{ssec:quan_eval}
\subsubsection{Overall Performance Under Anechoic Conditions}\label{sssec:results_anechoic}
\RI{While our methodology primarily targets complex reverberant environments, we begin with an anechoic signal reconstruction analysis to evaluate the frameworks where model-based assumptions are expected to hold. Table~\ref{tab:anechoic_analysis} shows \ac{stft-sdr} for the methods under study with one to four speakers. As shown in the table, the parametric methods (HO-DirAC and COMPASS) achieve the highest scores for a single speaker, as the anechoic single-source condition strictly satisfies the single-dominant-source assumption of HO-DirAC while providing an ideal low-complexity environment for COMPASS. However, their performance drops sharply with more speakers due to overlapping spatial components increasing the complexity for parameter estimation. Similarly, the spatial sparsity assumptions of \ac{pwd} \ac{cs} hold reasonably well across all mixtures, maintaining stable performance. Conversely, data-driven models show robustness to anechoic multi-speaker interference despite receiving no explicit anechoic training. Notably, in these low-uncertainty scenarios with 1-2 speakers, the diffusion framework underperforms the other data driven baselines. We attribute this to the low spatial ambiguity of the anechoic setting; under these conditions, a direct conditional mean-regression approach is highly optimal. In contrast, because the diffusion model employs an \ac{sde} sampling process designed to escape global averages, the continuous injection of stochastic noise introduces random structural fluctuations. This stochastic variance degrades both the fine spectral magnitude features and phase alignment, causing it to underperform. Reaffirming this, the \ac{fm} framework, which instead resolves the generative process via a deterministic \ac{ode} trajectory, retains a highly structured path that tracks the conditional mean velocity field without random spectral or phase degradations.}

\begin{table}[t]
\centering
\caption{\ac{stft-sdr} (dB) evaluated for all methods under anechoic conditions, for 1-4 speakers. Best results are bolded.}
\label{tab:anechoic_analysis}
\vspace{2mm}
\scriptsize
\begin{tabular}{@{}lcccc|c@{}}
\toprule
\textbf{Model} & \textbf{1 Spk} & \textbf{2 Spk} & \textbf{3 Spk} & \textbf{4 Spk} & \textbf{Avg.} \\
\midrule
PWD CS         & 12.90   & 14.34   &  12.38   & 10.94   & 12.63   \\
\RI{HO-DirAC}        & \RI{37.87}   & \RI{13.97}   &  \RI{7.89}   & \RI{5.22}   & \RI{15.65}  \\
\RI{COMPASS}        & \RI{\textbf{40.57}}   & \RI{15.95}   &  \RI{9.93}   & \RI{7.24}   & \RI{\textbf{17.82}}  \\
GRU-Uni                 & 11.50   & 9.04    &  7.06    & 5.63    & 8.31    \\
GRU-Bi                 & 12.30   & 9.89    &  7.89    & 6.40    & 9.12    \\
\ac{ctn}             & 14.05   & 11.77   &  9.76    & 8.60    & 10.97   \\
Diffusion               & 10.74   & 10.59   &  9.75    & 9.09    & 10.02   \\
Flow Matching           & 14.96 & \textbf{16.02} & \textbf{15.14} & \textbf{14.02} & 15.03 \\
\bottomrule
\end{tabular}
\end{table}

\begin{figure}
    % \vspace{-0.3cm}
    \centering
    \includegraphics[width=1\linewidth]{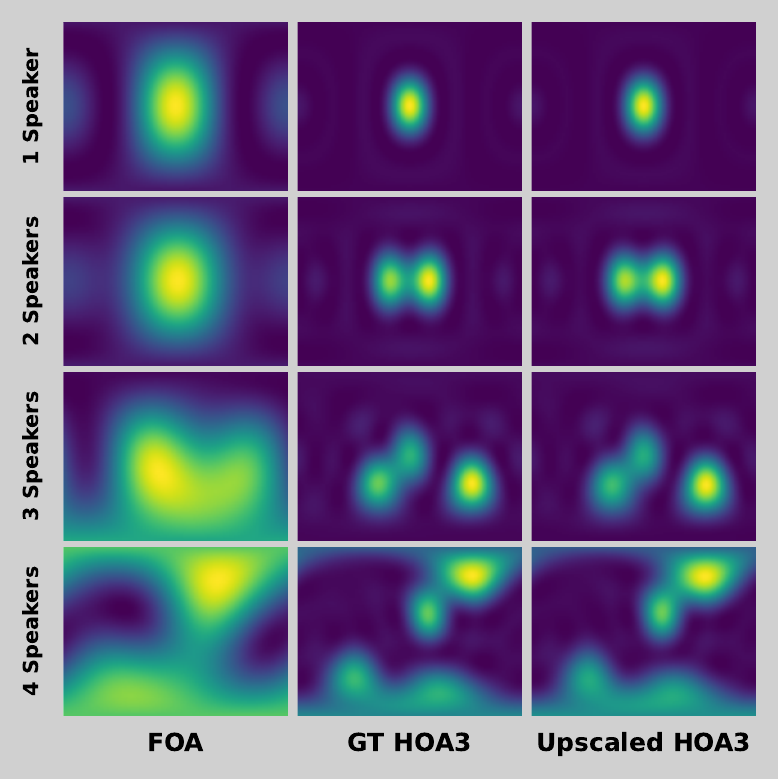}
    \caption{\small Directional energy plots over elevation (x-axis) and  azimuth (y-axis). Columns: \ac{foa}, ground truth \ac{hoa}3 and \ac{fm} upscaled \ac{hoa}3. Rows correspond to the number of active sources.}
    \label{fig:ambisonic_comparison}
\end{figure}

The anechoic setting also facilitates clear visualization of spatial directivity. Accordingly, we depict in  Fig.~\ref{fig:ambisonic_comparison} the directional energy plots of 3rd-order Ambisonics signals produced by \ac{fm} upscaler  with the ground truth for orders 1 and 3, across one to four active speakers. The results show a strong resemblance between the recovered \ac{hoa} and the reference energy patterns.

\subsubsection{Overall Performance Under Reverberant Conditions}\label{sssec:results_reverb_performance}
We now proceed to the primary focus of our experimental study: the evaluation of performance within complex reverberant environments. A high-level summary, representing the results averaged across all \ac{rt} and \ac{drr} subsets of the HARP dataset, is provided in Table~\ref{tab:harp_weighted_averaged_summary}.

\begin{table}[h]
\centering
\caption{\small \ac{stft-sdr} (dB) across \ac{rt} and \ac{drr} levels for the HARP evaluation set, detailed for 1--4 Speakers.}
\label{tab:harp_weighted_averaged_summary}
\vspace{2mm}
\scriptsize
\begin{tabular}{@{}lcccc|c@{}}
\toprule
\textbf{Model} & \textbf{1 Spk} & \textbf{2 Spk} & \textbf{3 Spk} & \textbf{4 Spk} & \textbf{Avg.} \\
\midrule
PWD CS         & 2.34 & 2.06 & 1.85 & 1.70 & 1.99 \\
\RI{HO-DirAC}       & \RI{3.09} & \RI{1.45} & \RI{0.71} & \RI{0.21} & \RI{1.36} \\
\RI{COMPASS}       & \RI{5.28} & \RI{3.35} & \RI{2.52} & \RI{1.97} & \RI{3.27} \\
GRU-Uni        & 5.46 & 3.82 & 3.03 & 2.60 & 3.73 \\
GRU-Bi         & 5.71 & 4.22 & 3.43 & 2.97 & 4.08 \\
\ac{ctn}    & 6.37 & 5.16 & 4.31 & 3.74 & 4.90 \\
Diffusion      & 6.26 & 4.98 & 4.04 & 3.49 & 4.69 \\
Flow Matching  & \textbf{8.79} & \textbf{7.97} & \textbf{7.08} & \textbf{6.26} & \textbf{7.53} \\
\bottomrule
\end{tabular}
\end{table}

Performance across all methods degrades in reverberant scenarios relative to anechoic conditions, which is expected given the highly stochastic nature of reverberant data. Specifically, \RI{the model-based methods exhibit significant degradation, as their underlying signal-model assumptions no longer hold in these complex acoustic environments}. Consequently, the remainder of our evaluation focuses primarily on data-driven methods.

The results indicate that the Diffusion framework achieves performance comparable to the \ac{ctn} baseline, both of which are significantly outperformed by the \ac{fm} framework across all speaker conditions. We attribute this performance gap to the fundamental differences in how these two generative frameworks model and navigate the data manifold. \RI{Standard signal reconstruction metrics, such as \ac{stft-sdr}, operate on point-wise structural alignment and heavily penalize any phase or realization mismatch. Because the deterministic \ac{ode} formulation of \ac{fm} tends to seek the conditional mean, it effectively synthesizes a smooth, expected representation of the target signal, maximizing average reconstruction metrics like \ac{stft-sdr}. 
In contrast, the diffusion model generates distinct, stochastic realizations of the true posterior. While this preserves natural acoustic textures, the resulting per-realization variance manifests as a penalty in strict point-wise distortion metrics, artificially widening the observed performance gap. Crucially, as we demonstrate in subsequent evaluations, this gap narrows significantly when shifting to spatial and perceptual criteria.}

% This hypothesis is further validated by the preliminary study \cite{milstein2025diffau} which demonstrated that in strictly anechoic settings, the Diffusion framework actually outperformed the \ac{ctn} baseline. This suggests that an unconstrained probability path is sufficient for signal reconstruction in the absence of complex interference. However, as the acoustic environment introduces stochastic reverberation patterns, this lack of constraint becomes a liability. These results confirm that while the Diffusion framework excels in simpler setups, the presence of reverberant stochasticity necessitates the more robust, linear trajectory provided by the \ac{fm} framework. \Amit{Should i remove this last justification? its only to further validate the hypothesis - not the compare results}

A similar trend is observed in the overall performance on the MOTUS dataset presented in Table~\ref{tab:motus_averaged}, characterized by a more substantial performance gap between the \ac{ctn} baseline and the diffusion framework, while the \ac{fm} framework consistently achieves the best performance.

% A similar trend is observed in the results for  the MOTUS dataset presented in Table~\ref{tab:motus_diffusion_subset}, despite the more restricted range of acoustic scenarios available in that recorded data.

% In particular, contrary to the results reported in \cite{milstein2025diffau}, where the diffusion  framework demonstrated superior performance in anechoic scenarios (when trained for such settings), our evaluations show it performing only comparably to the baseline\NirCmt{I am not sure we should discuss the results in \cite{milstein2025diffau} since this is the preliminary findings of this work and not a seperate work, at least that is how I see it. We should discuss next time we meet}. We attribute this discrepancy to the inherent stochasticity of the acoustic environments in the dataset.  Specifically, we hypothesize that the unconstrained probability path inherent in this framework, combined with the complex reverberation patterns in the data, limits the model's reconstruction accuracy. 
% %
% In contrast,  \ac{fm}  yields a significant performance gain over the baseline and diffusion framework across all acoustic scenarios. By constraining the probability path to be linear between the source and target distributions, \ac{fm} simplifies the mapping process. This linearity makes the model more robust to the underlying stochastic nature of reverberant data, enabling more consistent and accurate signal reconstruction.

\begin{table}[h]
\centering
\caption{\small \ac{stft-sdr} (dB) across \ac{rt} and \ac{drr} levels for the MOTUS  evaluation set, detailed for 1--4 Speakers.}
\label{tab:motus_averaged}
\vspace{2mm}
\scriptsize
\begin{tabular}{@{}lcccc|c@{}}
\toprule
\textbf{Model} & \textbf{1 Spk} & \textbf{2 Spk} & \textbf{3 Spk} & \textbf{4 Spk} & \textbf{Avg.} \\
\midrule
GRU-Uni        & 1.22 & 1.26 & 1.28 & 1.15 & 1.23 \\
GRU-Bi         & 2.09 & 2.15 & 2.21 & 2.02 & 2.12 \\
\ac{ctn}    & 3.47 & 3.25 & 3.12 & 2.77 & 3.15 \\
Diffusion      & 2.66 & 2.30 & 2.09 & 1.63 & 2.17 \\
Flow Matching  & \textbf{5.98} & \textbf{5.56} & \textbf{5.32} & \textbf{4.70} & \textbf{5.39} \\
\bottomrule
\end{tabular}
\end{table}

\subsubsection{The Effect of T60 and DRR on Performance}
We now present a comprehensive analysis using the HARP dataset, which encompasses a wide range of reverberant scenarios to more rigorously evaluate model performance across diverse acoustic conditions. Table~\ref{tab:harp_complete_matrix} details the performance of the various \ac{au} methods across the specific acoustic categories defined in Subsection~\ref{ssec:eval_data}.
\begin{table}[t]
\centering
\caption{STFT-SDR (dB) for the HARP evaluation set, divided by Reverberation (Rev.) and \ac{drr} levels for 1–4 speakers. Best results are bolded.}
\label{tab:harp_complete_matrix}
\vspace{2mm}
\scriptsize
\begin{tabular}{@{}lllcccc|c@{}}
\toprule
\textbf{Rev.} & \textbf{DRR} & \textbf{Model} & \textbf{1 Spk} & \textbf{2 Spk} & \textbf{3 Spk} & \textbf{4 Spk} & \textbf{Avg.} \\
\midrule

% --- LOW REVERB GROUP ---
\multirow{15}{*}{Low} 
& \multirow{5}{*}{Low} 
% & & PWD CS & 0.62 & 0.62 & 0.48 & 0.43 & 0.54 \\ &
& GRU-Uni & 2.16 & 1.69 & 1.27 & 1.20 & 1.61 \\
& & GRU-Bi & 2.26 & 1.86 & 1.39 & 1.33 & 1.74 \\
& & \ac{ctn} & 2.53 & 2.15 & 1.68 & 1.55 & 1.98 \\
& & Diffusion & 2.47 & 2.03 & 1.39 & 1.06 & 1.74 \\
& & Flow Matching & \textbf{4.29} & \textbf{3.96} & \textbf{3.31} & \textbf{2.96} & \textbf{3.63} \\
\cmidrule(lr){2-7} \cmidrule(l){8-8}
& \multirow{5}{*}{Med} 
% & & PWD CS & 2.01 & 1.81 & 1.63 & 1.55 & 1.75 \\ &
& GRU-Uni & 4.48 & 3.33 & 2.85 & 2.51 & 3.30 \\
& & GRU-Bi & 4.63 & 3.62 & 3.16 & 2.82 & 3.56 \\
& & \ac{ctn} & 5.07 & 4.24 & 3.75 & 3.36 & 4.11 \\
& & Diffusion & 5.55 & 4.63 & 3.95 & 3.46 & 4.40 \\
& & Flow Matching & \textbf{7.15} & \textbf{6.55} & \textbf{5.87} & \textbf{5.34} & \textbf{6.23} \\
\cmidrule(lr){2-7} \cmidrule(l){8-8}
& \multirow{5}{*}{High} 
% & & PWD CS & 5.40 & 4.72 & 4.09 & 3.61 & 4.45 \\ &
& GRU-Uni & 7.50 & 5.10 & 4.01 & 3.37 & 5.00 \\
& & GRU-Bi & 7.77 & 5.67 & 4.60 & 3.93 & 5.49 \\
& & \ac{ctn} & 8.73 & 7.03 & 5.95 & 5.07 & 6.70 \\
& & Diffusion & 8.72 & 7.15 & 6.11 & 5.20 & 6.80 \\
& & Flow Matching & \textbf{11.53} & \textbf{10.23} & \textbf{9.02} & \textbf{7.99} & \textbf{9.69} \\
\midrule
% --- MEDIUM REVERB GROUP ---
\multirow{15}{*}{Med} 
& \multirow{5}{*}{Low} 
% & & PWD CS & 0.64 & 0.74 & 0.56 & 0.48 & 0.60 \\ & NOT WITH <10 drr
& GRU-Uni & 2.10 & 1.69 & 1.32 & 1.14 & 1.54 \\
& & GRU-Bi & 2.24 & 1.87 & 1.50 & 1.31 & 1.71 \\
& & \ac{ctn} & 2.53 & 2.20 & 1.84 & 1.56 & 2.03 \\
& & Diffusion & 2.35 & 2.00 & 1.47 & 1.10 & 1.73 \\
& & Flow Matching & \textbf{3.45} & \textbf{3.35} & \textbf{2.87} & \textbf{2.61} & \textbf{3.07} \\
\cmidrule(lr){2-7} \cmidrule(l){8-8}
& \multirow{5}{*}{Med} 
% & & PWD CS & 1.92 & 1.64 & 1.45 & 1.33 & 1.59 \\ & NOT WITH <10 drr
& GRU-Uni & 4.25 & 3.10 & 2.63 & 2.25 & 3.07 \\
& & GRU-Bi & 4.45 & 3.46 & 3.00 & 2.60 & 3.39 \\
& & \ac{ctn} & 5.20 & 4.32 & 3.74 & 3.27 & 4.13 \\
& & Diffusion & 5.61 & 4.69 & 4.05 & 3.37 & 4.43 \\
& & Flow Matching & \textbf{7.31} & \textbf{6.64} & \textbf{6.06} & \textbf{5.23} & \textbf{6.31} \\
\cmidrule(lr){2-7} \cmidrule(l){8-8}
& \multirow{5}{*}{High} 
% & & PWD CS & 4.18 & 3.52 & 3.13 & 3.04 & 3.47 \\ & NOT WITH <10 drr
& GRU-Uni & 7.82 &  5.26 & 4.04 & 3.74 & 5.22 \\
& & GRU-Bi & 8.10 & 5.89 & 4.61 & 4.30 & 5.73 \\
& & \ac{ctn} & 9.58 & 7.71 & 6.28 & 5.90 & 7.37 \\
& & Diffusion & 9.00 & 7.56 & 6.47 & 6.07 & 7.28 \\
& & Flow Matching & \textbf{11.77} & \textbf{10.59} & \textbf{9.38} & \textbf{8.92} & \textbf{10.17} \\
\midrule
% --- HIGH REVERB GROUP ---
\multirow{15}{*}{High} 
& \multirow{5}{*}{Low} 
% & & PWD CS & 0.96 & 0.81 & 0.72 & 0.66 & 0.79 \\ & NOT WITH <10 drr
& GRU-Uni & 2.40 & 1.73 & 1.52 & 1.33 & 1.69 \\
& & GRU-Bi & 2.59 & 1.93 & 1.72 & 1.53 & 1.89 \\
& & \ac{ctn} & 3.13 & 2.54 & 2.16 & 1.93 & 2.44 \\
& & Diffusion & 3.02 & 2.29 & 1.86 & 1.47 & 2.16 \\
& & Flow Matching & \textbf{4.89} & \textbf{4.10} & \textbf{3.70} & \textbf{3.34} & \textbf{4.01} \\
\cmidrule(lr){2-7} \cmidrule(l){8-8}
& \multirow{5}{*}{Med} 
% & & PWD CS & 1.85 & 1.57 & 1.44 & 1.28 & 1.53 \\ & NOT WITH <10 drr
& GRU-Uni & 4.17 & 3.06 & 2.49 & 2.08 & 2.97 \\
& & GRU-Bi & 4.41 & 3.44 & 2.82 & 2.41 & 3.29 \\
& & \ac{ctn} & 5.18 & 4.30 & 3.55 & 3.18 & 4.05 \\
& & Diffusion & 5.17 & 4.39 & 3.48 & 3.11 & 4.04 \\
& & Flow Matching & \textbf{7.28} & \textbf{6.49} & \textbf{5.77} & \textbf{5.24} & \textbf{6.20} \\
\cmidrule(lr){2-7} \cmidrule(l){8-8}
& \multirow{5}{*}{High} 
% & & PWD CS & 3.41 & 2.94 & 2.61 & 2.39 & 2.84 \\ & NOT WITH <10 drr
& GRU-Uni & 6.88 & 4.62 & 3.88 & 3.08 & 4.63 \\
& & GRU-Bi & 7.17 & 5.21 & 4.45 & 3.57 & 5.10 \\
& & \ac{ctn} & 8.25 & 6.79 & 5.88 & 4.89 & 6.45 \\
& & Diffusion & 7.35 & 6.49 & 5.72 & 4.94 & 6.13 \\
& & Flow Matching & \textbf{9.53} & \textbf{8.85} & \textbf{8.13} & \textbf{7.29} & \textbf{8.45} \\
\bottomrule
\end{tabular}
\end{table}

\begin{figure*}
    \centering
    \includegraphics[width=1\linewidth]{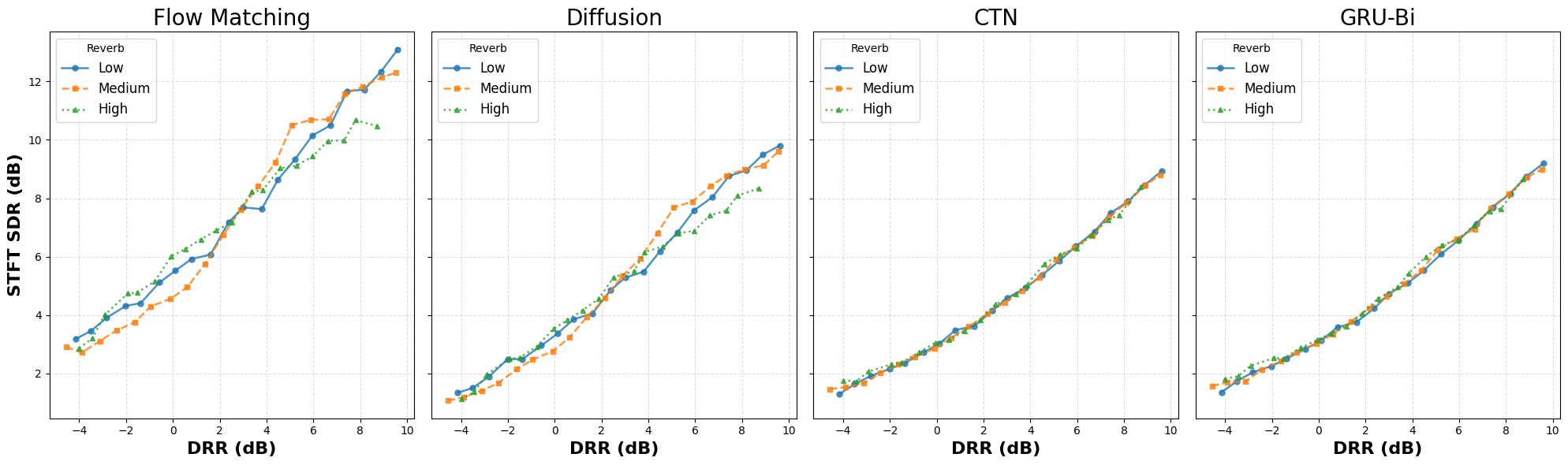}
    \caption{Comparative analysis of \ac{stft-sdr} (dB) performance across models as a function of \ac{drr}, categorized by \ac{rt} levels.}
    \label{fig:sdr_drr_plots}
\end{figure*}

The results indicate that \ac{drr} exerts a significantly more profound influence on performance than either \ac{rt} or the number of speakers. To further illustrate this behavior, Fig.~\ref{fig:sdr_drr_plots} depicts performance across a continuous range of \ac{drr} values, grouped by the defined \ac{rt} subsets for single-speaker scenarios. Notably, the performance curves for the various \ac{rt} subsets exhibit nearly identical trajectories; this reinforces the premise that reconstruction quality is primarily driven by \ac{drr}, maintaining a consistent trend across model architectures regardless of the specific reverberation category. These trends indicate that while the complexity of the 
acoustic scene increases with more speakers, the primary bottleneck for accurate signal reconstruction is the \ac{drr} energy 
rather than the temporal decay characteristics of the room.

% The results in Table~\ref{tab:harp_complete_matrix} indicate that  the \ac{drr} exerts a far greater influence on performance than the \ac{rt} or the number of speakers. This is further evidenced in Table~\ref{tab:harp_summary}, where the standard deviation across different 
% reverberation times ($\sigma_{T_{60}}$) remains notably low, often below 1~dB for Low and Medium \ac{drr} conditions. These trends indicate that while the complexity of the 
% acoustic scene increases with more speakers, the primary bottleneck for accurate signal reconstruction is the \ac{drr} energy 
% rather than the temporal decay characteristics of the room.
% %
% To further verify this trend, we evaluate the models on single-speaker scenarios in Fig.~\ref{fig:combined_sdr_plots} across a continuous range of \ac{drr} values, grouped by our defined \ac{rt} 
% subsets.   Fig.~\ref{fig:t60_drr} illustrates the distribution of these acoustic parameters within each subset, while Fig.~\ref{fig:sdr_drr} highlights the resulting performance across the \acp{drr}. Notably, the performance curves for the different \ac{rt} subsets exhibit a highly similar trajectory; this reinforces the premise that the reconstruction quality is primarily driven by the \ac{drr}, maintaining a consistent trend regardless of the specific reverberation category across model architectures.

% \begin{figure*}[t]
%     \centering
%     \begin{subfigure}{0.48\linewidth}
%         \centering
%         \includegraphics[width=\linewidth]{figs/T60_DRR_plot.png}
%         \caption{} 
%         \label{fig:t60_drr} % <--- Reference this for (a)
%     \end{subfigure}
%     \hfill 
%     \begin{subfigure}{0.48\linewidth}
%         \centering
%         \includegraphics[width=\linewidth]{figs/SDR_DRR_plot.png}
%         \caption{} 
%         \label{fig:sdr_drr} % <--- Reference this for (b)
%     \end{subfigure}
%     \caption{\ac{stft-sdr} analysis: (a) Distribution of \ac{rt} values categorized by reverberation intensity (low, medium, and high) across the \ac{drr} range; and (b) corresponding \ac{stft-sdr} performance.\Amit{Rethink how to present it - too clustered}}
%     \label{fig:combined_sdr_plots} % <--- Reference this for the whole thing
% \end{figure*}

\subsubsection{Frequency-Dependent Coherence}
\begin{figure}
    \centering
    \includegraphics[width=1\linewidth]{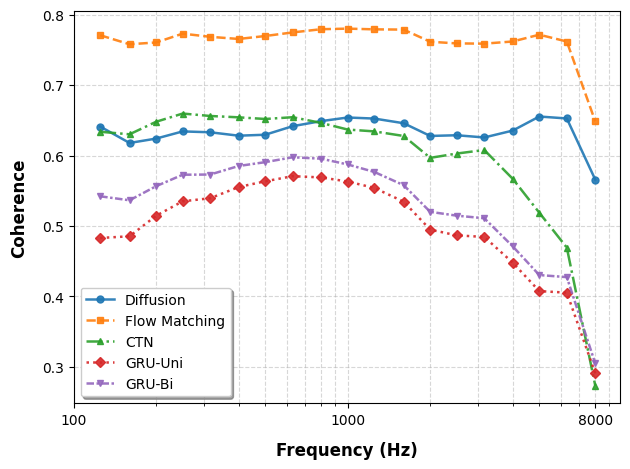}
    \caption{Average coherence $\bar{\gamma}_f^2$ versus frequency.}
    \label{fig:coherence}
\end{figure}

We proceed to evaluate the performance across the frequency spectrum for all methods, with the coherence averaged over channels, as defined in \eqref{eqn:coherence}, which is illustrated in Fig.~\ref{fig:coherence}. Notably, 
the generative approaches maintain consistent performance across the entire spectrum, whereas 
the discriminative ones degrade significantly at higher frequencies. This decay likely stems from the use of deterministic regression losses, which tend to average out high-frequency phase details to minimize point-wise error under uncertainty. In contrast, generative models leverage learned data distributions to synthesize realistic high-frequency components, maintaining structural coherence.

\subsubsection{Spatial Performance Analysis} \label{sssec:results_spatial_performance}
Lastly, we evaluate spatial performance of the \ac{au} methods. We evaluate the spatial accuracy, prioritizing a per-\ac{drr} analysis to highlight the governing factors of the spatial reconstruction.
\begin{table}[t]
\centering
\caption{Spatial performance in terms of $\Delta$DI and $\Delta\sigma$ across \ac{drr} levels. Lower values indicate better reconstruction of the spatial field. Best results are bolded.}
\label{tab:spatial_feat_comparison}
\scriptsize
\begin{tabular}{llcc}
\toprule
\textbf{DRR} & \textbf{Model} & \textbf{$\Delta$ DI (dB) $\downarrow$} & \textbf{$\Delta\sigma$ ($^\circ$) $\downarrow$} \\ \midrule
\multirow{6}{*}{Low}    & Input (FOA)           & 5.25          & 4.68 \\
& GRU-Uni (HOA-3)       & 1.16          & 3.33 \\ 
& GRU-Bi (HOA-3)        & 1.23          & 3.14 \\ 
& \ac{ctn} (HOA-3)   & 1.46          & 2.61 \\ 
& Diffusion (HOA-3)     & \textbf{0.23} & 1.51 \\
& Flow Matching (HOA-3) & 0.90          & \textbf{0.45} \\ \addlinespace

\multirow{6}{*}{Med}    & Input (FOA)           & 5.62          & 2.89 \\
& GRU-Uni (HOA-3)       & 0.71          & 2.16 \\
& GRU-Bi (HOA-3)        & 0.74          & 1.96 \\
& \ac{ctn} (HOA-3)   & 0.76          & 1.68 \\ 
& Diffusion (HOA-3)     & \textbf{0.11} & 1.22 \\
& Flow Matching (HOA-3) & 0.40          & \textbf{0.83} \\ \addlinespace

\multirow{6}{*}{High}   & Input (FOA)           & 5.86          & 1.85 \\
& GRU-Uni (HOA-3)       & 0.28          & 1.52 \\
& GRU-Bi (HOA-3)        & 0.30          & 1.36 \\
& \ac{ctn} (HOA-3)   & 0.19          &  1.11 \\
& Diffusion (HOA-3)     & \textbf{0.06} & 1.06 \\
& Flow Matching (HOA-3) & 0.14          & \textbf{0.67} \\ \bottomrule
\end{tabular}
\end{table}

Table \ref{tab:spatial_feat_comparison} presents the spatial performance of each approach relative to the ground-truth \ac{hoa}3 on reverberant single-speaker scenarios, with \ac{foa} performance included as a reference baseline. As illustrated, the generative approaches exhibit spatial characteristics that more closely align with the ground truth than those of the baseline models. We attribute this advantage to the fundamental differences in their optimization objectives. While the discriminative baseline minimizes a global reconstruction error, it inherently suffers from regression to the mean, resulting in spatially blurred outputs where fine-grained directional information is lost. In contrast, by modeling the score function or the probability flow of the \ac{hoa} data, the generative frameworks successfully reconstruct signals that adhere to the true data manifold, thereby preserving delicate spatial characteristics.

\RI{Moreover, a distinct trade-off between the two generative paradigms emerges in Table \ref{tab:spatial_feat_comparison}: while the \ac{sgm} framework achieves a lower $\Delta\text{DI}$ compared to \ac{fm}, the trend for $\Delta\sigma$ is reversed. This phenomenon stems from a subtle distinction in how these two methods sample the generative space. Unlike the global mean-regression of the discriminative baseline, the deterministic \ac{ode} trajectories of \ac{fm} track the local conditional mean when confronted with intersecting generation paths. This condition becomes increasingly prevalent in highly reverberant data, introducing localized spatial blurring that manifests as an increased $\Delta\text{DI}$. By comparison, \ac{sgm} acts as a stochastic, unbiased sampler during inference; its \ac{sde} formulation continuously injects random noise, enabling the generative process to ``escape'' these local mean-seeking intersections and converge toward a single, sharp realization of the true posterior. This interpretation is further validated by the fact that the gap in $\Delta\text{DI}$ widens as the DRR decreases, suggesting a higher density of trajectory intersections in heavily reverberant environments for \ac{fm} path.}

\section{Listening Test I - Acoustic Scenario Comparison}\label{ssec:listening_exp_I}
To complement the previous quantitative evaluations, listening tests were conducted\footnote{Audio samples comparing all evaluated methods are available online at \url{https://amitmils.github.io/GenAU}}. While metrics reported in Section \ref{sec:emp_eval} are essential for quantitatively assessing \ac{au} in terms of signal reconstruction and directivity, they do not always capture the full complexity of human auditory perception. This study serves as an investigation into the subjective quality of \RI{our proposed method,} %synthesized \ac{hoa} signals,
exploring whether the observed technical trends \RI{across acoustic scenarios} are reflected in listener preferences. To that aim, we first describe the listening test setup and methodology in Subsections~\ref{ssec:listening_setup_I} and \ref{ssec:listening_methodology_I}, respectively, while Subsection~\ref{ssec:listening_results_I} provides the results and a discussion of the findings.

\subsection{Setup}\label{ssec:listening_setup_I}
All signals were generated by convolving 6-second dry speech samples from the WSJ0 corpus with simulated \acp{sh-rir} from the HARP codebase. Notably, both the dry signals and the specific room configurations used for the \ac{sh-rir} simulations were withheld from the training phase to ensure data integrity.

The evaluation comprised two distinct listening tests, totaling seven acoustic scenarios. This design was implemented to determine if subjective perception aligns with the quantitative trends observed in Subsection~\ref{ssec:quan_eval}. To isolate the acoustic environment as the primary variable, the speaker identity, utterance, and source position were kept constant within each test group:
\begin{enumerate}
    \item \textbf{\ac{drr} Variation Group:} Four scenarios with a fixed \ac{rt} of 500~ms were evaluated to determine the impact of \ac{drr} variation across different room scales. A single male speaker was selected, positioned at  azimuth of $-60^\circ$ and an elevation of $90^\circ$:
    \begin{description}
        \item[Scenario A:] $6.8 \times 5.0 \times 2.5$~m ($-2.5$~dB \ac{drr})
        \item[Scenario B:] $6.2 \times 8.0 \times 2.6$~m ($0.0$~dB \ac{drr})
        \item[Scenario C:] $12.0 \times 9.0 \times 2.2$~m ($3.5$~dB \ac{drr})
        \item[Scenario D:] $17.6 \times 21.2 \times 4.1$~m ($9.5$~dB \ac{drr})
    \end{description}

    \item \textbf{Reverberation Variation Group:} Three scenarios were defined with a constant \ac{drr} of 6~dB to investigate the independent effect of decay time on perceived quality. For this group, a single female speaker was used, positioned at an azimuth of $60^\circ$ and an elevation of $90^\circ$:
    \begin{description}
        \item[Scenario E:] $9.2 \times 13.4 \times 2.2$~m (\ac{rt} = 270~ms)
        \item[Scenario F:] $16.0 \times 15.0 \times 2.5$~m (\ac{rt} = 450~ms)
        \item[Scenario G:] $18.0 \times 15.5 \times 3.2$~m (\ac{rt} = 750~ms)
    \end{description}
\end{enumerate}

\begin{figure*}[t]
    \centering
    \begin{subfigure}{0.48\linewidth}
        \centering
        \includegraphics[width=\linewidth]{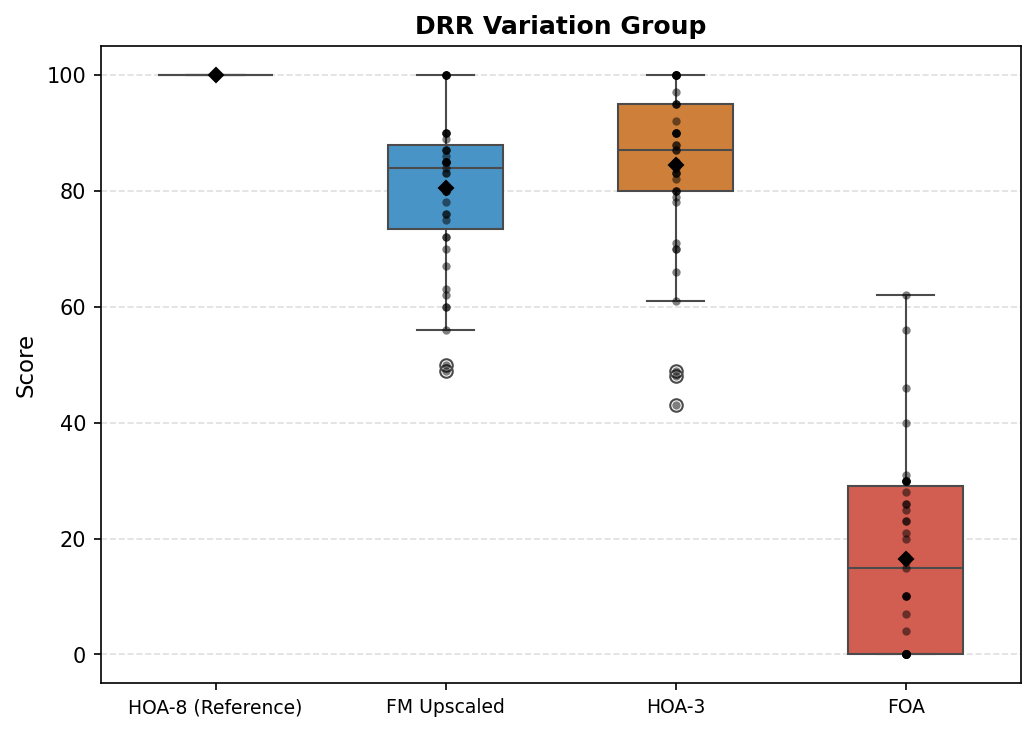}
        \caption{} 
        \label{fig:mushra_drr_sweep} % <--- Reference this for (a)
    \end{subfigure}
    \hfill 
    \begin{subfigure}{0.48\linewidth}
        \centering
        \includegraphics[width=\linewidth]{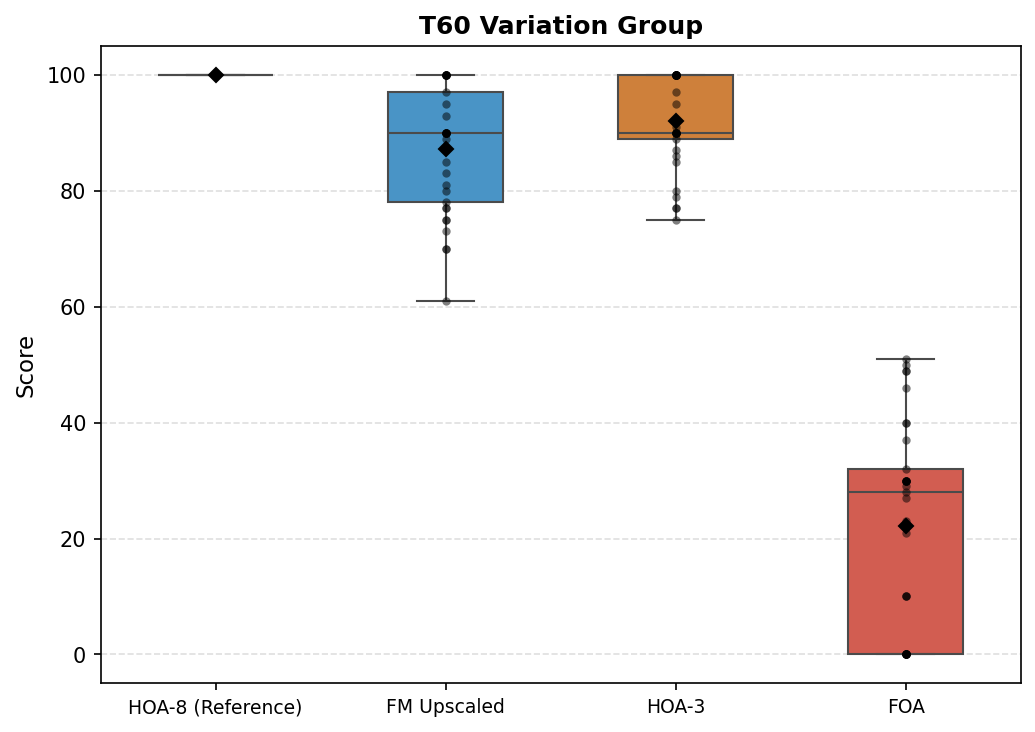}
        \caption{} 
        \label{fig:mushra_rt_sweep} % <--- Reference this for (b)
    \end{subfigure}
    \caption{MUSHRA Results: (a) scores for \ac{drr} variation group; and (b) scores for \ac{rt} variation group}
    \label{fig:combined_mushra_plots} % 
\end{figure*}

For each scenario, four distinct stimuli were prepared. First, \acp{sh-rir} were generated at three Ambisonics orders: 8th, 3rd, and 1st (\ac{foa}). These were convolved with the dry signals to produce the corresponding Ambisonics signals. The fourth stimulus was created by upscaling the \ac{foa} signal using the proposed \ac{fm} framework.
Lastly, all signals were rendered binaurally using the least-squares method~\cite{Avni2013spatial} with Neumann KU100 HRTFs and normalized to $-6$~LUFS to ensure perceptual loudness equality across all conditions. To focus the assessment strictly on static spatial accuracy, no head-tracking was employed during the evaluation.

\subsection{Methodology}\label{ssec:listening_methodology_I}

The subjective evaluation followed a \ac{mushra} protocol~\cite{series2014method}. Participants were tasked with rating the \textit{overall spatial fidelity} of four stimuli compared to a known reference: $(i)$ a hidden 8th-order reference, $(ii)$ a \ac{foa} anchor, $(iii)$ an upscaled \ac{hoa}3 signal using  the \ac{fm} framework, and $(iv)$ the ground-truth \ac{hoa}3. 

The evaluation was conducted by 11 listeners, each of whom participated in a brief training session to become familiar with the range of stimuli and the interface. To mitigate potential order bias, the acoustic scenarios were presented in a randomized sequence.

The participants provided ratings on a continuous scale from 0 to 100, where a score of 100 represents no audible difference from the reference, where lower scores indicate larger differences.

\subsection{Results}\label{ssec:listening_results_I}
The results of both experiments are illustrated in Fig. \ref{fig:combined_mushra_plots}. 

\subsubsection{\ac{drr} Variation Group}
A two-way repeated-measures \ac{anova} was conducted to evaluate the impact of the processing method and \ac{drr} levels on subjective quality ratings.
The analysis of the main effect of the processing method first revealed that Mauchly’s test of sphericity was violated ($\chi^2(5) = 15.00, p = 0.011$); consequently, Greenhouse-Geisser corrections were applied to the degrees of freedom. The results indicate a statistically significant main effect of the method ($F(3, 30) = 149.17, p < .001, \eta_p^2 = 0.94$), suggesting that the processing method significantly influences perceived quality regardless of the acoustic environment.
%\NirCmt{I don't understand any of the notations here. Are they conventional is spatial audio? should they be defined? I leave the review of this section to Boaz}.

In contrast, the main effect of the \ac{drr} level was not statistically significant ($F(3, 30) = 0.35, p = 0.787, \eta_p^2 = 0.03$), indicating that variations in the \ac{drr} within the tested range did not significantly alter the subjective scores. Furthermore, no significant interaction between the processing method and \ac{drr} was observed ($F(9, 72) = 1.86, p = 0.190, \eta_G^2 = 0.06$). This suggests that the performance of the evaluated methods remained robust and consistent across the different \ac{drr} scenarios.

Finally, post-hoc pairwise comparisons were performed to evaluate the proposed system against the ground-truth \ac{hoa}3 across all \ac{drr} conditions; no significant differences were found in any of the tested scenarios ($p > 0.05$). 
Despite the lack of statistical significance, this per-scenario analysis revealed a clear perceptual trend: the gap between the upscaled output and the ground truth narrowed consistently as the \ac{drr} increased. At the lowest \ac{drr} condition ($\ac{drr} = -2.5$~dB), a mean score difference of $9.4$ was observed with a large effect size ($g = 0.82$), favoring the ground truth. As the \ac{drr} 
increased to $0.0$~dB and $3.5$~dB, the mean score difference reduced to $4.3$ ($g = 0.41$) and $1.1$ ($g = 0.07$), respectively. Finally, at $9.5$~dB, the difference reached $-2.09$ ($g = -0.11$), indicating that the perceptual gap had effectively closed and slightly reversed in favor of the upscaled output. This convergence in performance across higher \ac{drr} scenarios directly correlates with the objective trends observed in Subsection~\ref{ssec:quan_eval}.

\subsubsection{\ac{rt} Variation Group}

A two-way repeated-measures ANOVA was also conducted to analyze the influence of the processing method and \ac{rt} levels on subjective quality. 
Similar to the \ac{drr} group, Mauchly’s test indicated that the assumption of sphericity was violated for the main effect of the method ($\chi^{2}(5) = 23.59, p < .001$), requiring Greenhouse-Geisser corrections. The analysis revealed a statistically significant main effect for the processing method ($F(3, 30) = 136.76, p < .001, \eta_{p}^{2} = 0.93$). Conversely, no statistically significant main effect was found for the $T_{60}$ level ($F(2, 20) = 0.66, p = 0.526, \eta_{p}^{2} = 0.06$), and the interaction between method and $T_{60}$ was likewise non-significant ($F(6, 60) = 1.23, p = 0.305, \eta_{G}^{2} = 0.03$).

Post-hoc pairwise comparisons between the proposed system and the \ac{hoa}3 ground truth further elucidated the relationship between decay time and perceived quality. While the global difference was not statistically significant ($p = 0.298, g = 0.64$), the per-scenario analysis confirmed that no significant differences were found in any individual scenario ($p > 0.05$). Despite this, a clear perceptual trend emerged: the performance gap widened as the reverberation time increased. Specifically, at the lowest reverberation level ($T_{60} = 0.27$~s), the mean difference score was $1.64$ with a small effect size ($g = 0.16$), indicating the upscaled signal was perceptually equal to the ground truth. However, as the $T_{60}$ increased to $0.45$~s and $0.75$~s, the mean difference score grew to $6.0$ ($g = 0.54$) and $6.8$ ($g = 0.82$), respectively. This trend suggests that while the proposed method performs excellently in less reverberant conditions, the perceptual difference from the ground truth becomes more pronounced as the effect of reverberation increases (although not statistically significant), a result that aligns with the objective trends observed in Subsection~\ref{ssec:quan_eval}.

\subsubsection{Summary}
While the \ac{fm} framework outperformed the baseline in the quantitative analysis in Section \ref{sec:emp_eval}, the listening test results indicate a remaining (although statistically insignificant) perceptual gap between the upscaled output and the ground truth \ac{hoa}3 in highly reverberant scenarios. Closing this gap remains a primary objective for future work. Nevertheless, we believe these results demonstrate that a generative approach to \ac{au} is a highly promising direction, providing a robust theoretical framework for addressing what is inherently a severely underdetermined inverse problem.

\RI{
\section{Listening Test II - \ac{au} Methods Comparison}\label{ssec:listening_exp_II}
We conduct an additional listening test to compare different \ac{au} methods, spanning model-based, generative, and discriminative data-driven approaches. To this end, Subsections~\ref{ssec:listening_setup_II} and \ref{ssec:listening_methodology_II} describe the setup and methodology, respectively, while Subsection~\ref{ssec:listening_results_II} presents the results and a discussion of the findings.
\subsection{Setup}\label{ssec:listening_setup_II}
We compare single- and multi-speaker settings using the same room as in \emph{Scenario~F} (\ac{rt}$=450\text{ms}$,  \ac{drr}$=6\text{dB}$)  from the preceding section. The first scenario evaluated a single speaker positioned at $(\theta, \phi) = (90^\circ, 20^\circ)$, whereas the second evaluated a multi-speaker scene with three concurrent speakers located at $(90^\circ, 30^\circ)$, $(90^\circ, -60^\circ)$, and $(90^\circ, 90^\circ)$, respectively. All signals were generated as described in Subsection~\ref{ssec:listening_setup_I}.
\begin{figure}
    \centering
    \includegraphics[width=1\linewidth]{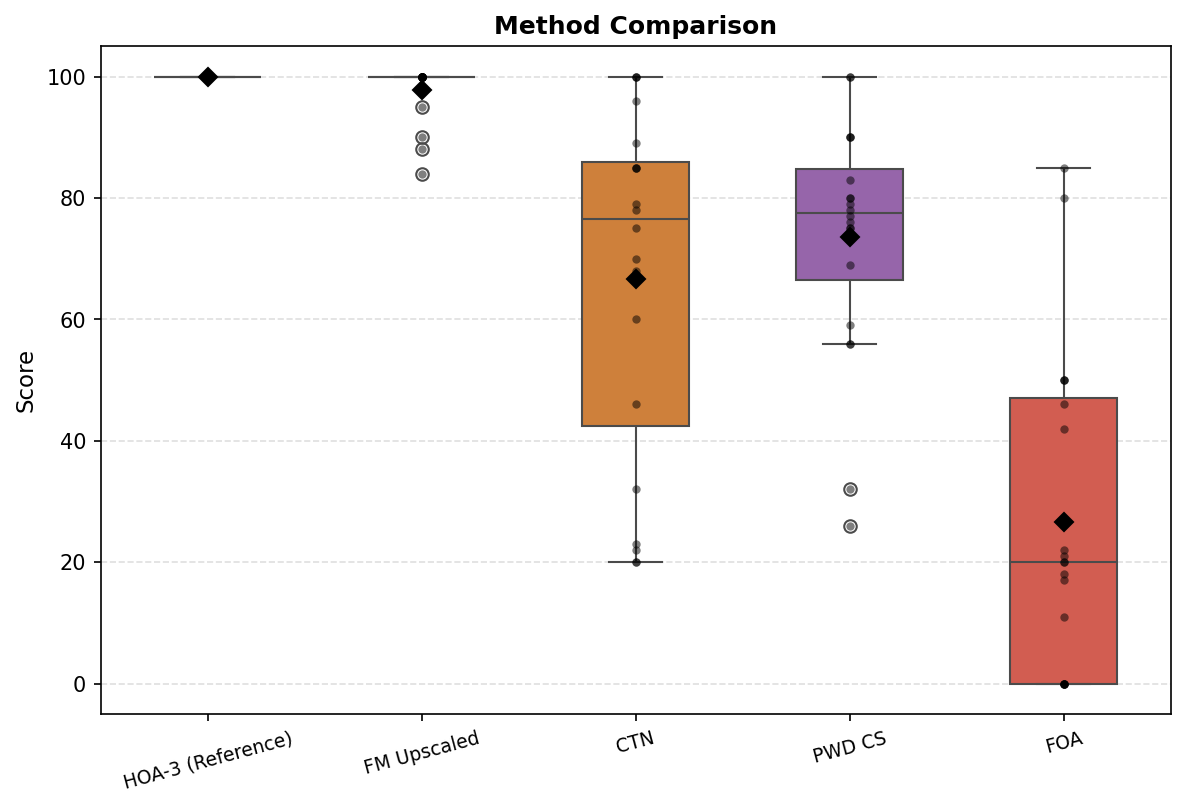}
    \caption{MUSHRA results for \ac{au} method comparison}
    \label{fig:mushra_results_II}
\end{figure}
\subsection{Methodology}\label{ssec:listening_methodology_II}
We followed the same \ac{mushra} methodology as in Subsection~\ref{ssec:listening_methodology_I}, with the exception that five stimuli were presented: $(i)$~a hidden \ac{hoa}3 reference, $(ii)$~a \ac{foa} anchor, and 3rd-order \ac{hoa} signals upscaled from \ac{foa} using $(iii)$~the proposed \ac{fm} framework, $(iv)$~\ac{ctn} (a deterministic baseline), and $(v)$~PWD CS (a model-based baseline). With the aim of focusing on between-method comparison while keeping the number of stimuli small, the 8th-order Ambisonics reference was omitted. The evaluation was conducted by 11~listeners experienced in spatial audio testing.
\subsection{Results}\label{ssec:listening_results_II}
Results of the method comparison experiment are presented in Fig.~\ref{fig:mushra_results_II}.
A two-way repeated-measures \ac{anova} was conducted to evaluate the impact of the processing method. Mauchly’s test of sphericity was violated ($\chi^2(9) = 37.11, p < .001$); consequently, Greenhouse-Geisser corrections were applied to the degrees of freedom. The results indicate a statistically significant main effect of the method ($F(1.62, 14.56) = 25.02, p < .001, \eta_p^2 = 0.94$), suggesting that the processing method significantly influences perceived quality. As can be seen, \ac{ctn} scored significantly lower than the proposed method and performed comparably to the model-based baseline, despite outperforming the latter in signal fidelity. Listener feedback indicated that this subjective degradation was driven by intermittent clicking artifacts in the \ac{ctn} output, which were noticed by a subset of participants. 
Post-hoc pairwise comparisons with Bonferroni correction between the proposed system and other methods revealed no significant difference compared to ground truth \ac{hoa}3 ($p = 0.837$, $g = -0.83$). The comparison to \ac{ctn} showed a trend toward significance but did not cross the standard threshold 
($p = 0.051$, $g = -1.61$), whereas the difference compared to PWD CS was statistically significant ($p = 0.013$, $g = -1.73$). These findings demonstrate that the \ac{fm} method matches the perceptual quality of ground-truth \ac{hoa}3, while the competing approaches produced noticeable artifacts.}

%----------------------------------------------------------------------------------------
%	CONCLUSIONS
%----------------------------------------------------------------------------------------
% \vspace{-0.2cm}
\section{Conclusions}\label{sec:conclusions}

In this work, we proposed a novel generative framework for \ac{au}, formulating the spatial upscaling of reverberant speech as a principled inverse problem. By learning the underlying data manifold, the proposed method provides a robust solution for synthesizing high-order spatial audio across diverse acoustic conditions. Our comprehensive evaluation across varied scenarios demonstrates the clear benefits of this approach for \ac{au}. Furthermore, while not perceptually identical to the \ac{hoa}3 ground truth in all scenarios, the performance gap was not statistically significant. 

\RI{Our framework requires a larger parameter footprint and higher computational complexity than deterministic baselines, which may limit its immediate real-time utility. However, it successfully demonstrates the strong potential of generative paradigms to capture complex spatial distributions with high fidelity. Accordingly, future work will focus on refining the generative process to preserve nuanced spatial cues, alongside architectural optimization and model reduction to facilitate real-time deployment. We also intend to explore latent-space formulations to efficiently scale to higher sampling rates.}

% Our evaluation demonstrates that the framework achieves high perceptual transparency and consistently small effect sizes, nearing the performance of \ac{hoa}3 ground truth. 
% Future research will focus on refining the generative process to further preserve nuanced spatial cues, ultimately closing the remaining perceptual gap in high-order reconstructions.

% In this work, we proposed a novel generative framework for \ac{au}, providing a mathematically grounded approach to the inverse problem of spatial upscaling through generative-based modeling. 

% Our comprehensive evaluation of reverberant speech across diverse acoustic scenarios demonstrates that while the proposed method is not yet perceptually equivalent to the \ac{hoa}3 ground truth in a strictly statistical sense, it achieves a high degree of perceptual transparency with remarkably small effect sizes across most conditions.  

% \textcolor{red}{Amit - I know that nobody really reads conclusions, but still we need it to reflect the paper, and kind of summarize it in one paragraph. The current conclusions section only discusses the experimental study (with the exception of the first sentence). Please rewrite it to reflect the entire paper (and keep it at most 4 sentences as indeed no one reads it)}
%----------------------------------------------------------------------------------------
%	REFERENCES
%----------------------------------------------------------------------------------------
\bibliographystyle{IEEEtran}
\bibliography{IEEEabrv,refs}

\begin{IEEEbiography}[{\includegraphics[width=1in,height=1.25in,clip,keepaspectratio]{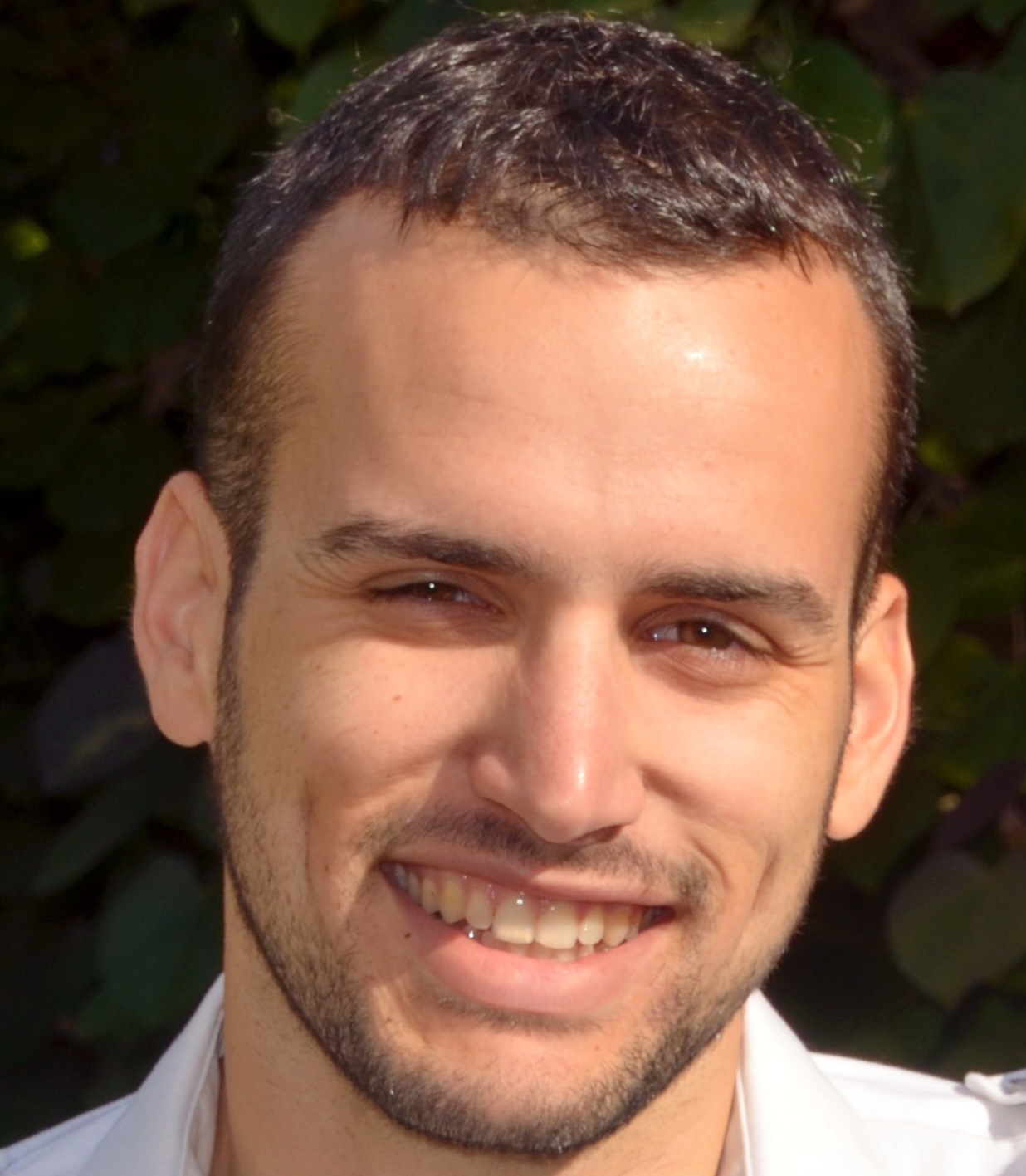}}]{Amit Milstein}  received his B.Sc. (2021) and M.Sc. (2024) degrees in electrical and computer engineering from Ben-Gurion University of the Negev, where he is currently pursuing his Ph.D.
\end{IEEEbiography}

\begin{IEEEbiography}[{\includegraphics[width=1in,height=1.25in,clip,keepaspectratio]{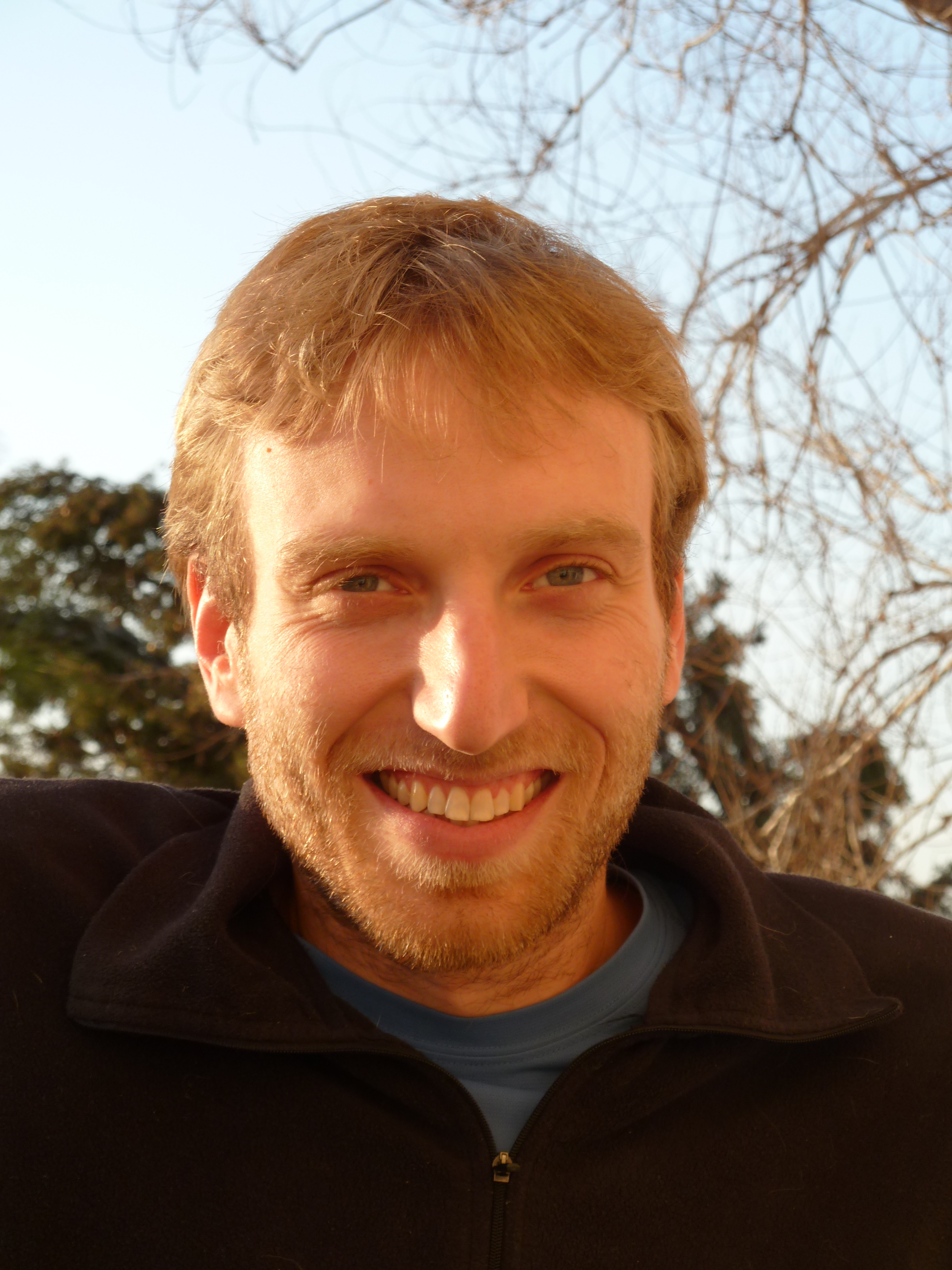}}]{Nir Shlezinger} (M’17-SM’23)  is an associate professor in the School of Electrical and Computer Engineering at Ben-Gurion University, Israel. He received his B.Sc., M.Sc., and Ph.D. degrees in 2011, 2013, and 2017, respectively, from Ben-Gurion University, Israel, all in electrical and computer engineering. From 2017 to 2019, he was a postdoctoral researcher at the Technion, and from 2019 to 2020, he was a postdoctoral researcher at the Weizmann Institute of Science, where he was awarded the FGS Prize for his research achievements. He is the recipient of the 2024 IEEE ComSoc Fred W. Ellersick Award, the 2025 IEEE ComSoc Marconi Prize, and the 2024 Krill Prize for outstanding young researchers.  His research interests include communications, information theory, signal processing, and machine learning.
\end{IEEEbiography}

\begin{IEEEbiography}[{\includegraphics[width=1in,height=1.25in,clip,keepaspectratio]{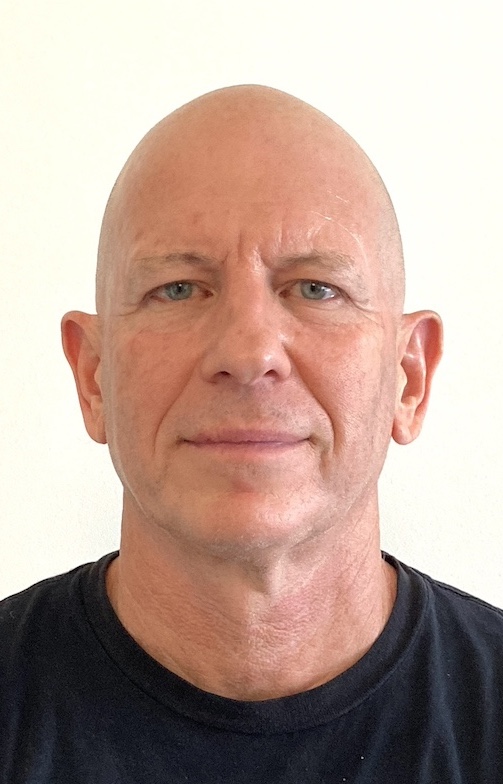}}]{Boaz Rafaely} (Fellow, IEEE) received the B.Sc. degree (cum laude) in electrical engineering from Ben-Gurion University, Beer-Sheva, Israel, in 1986; the M.Sc. degree in biomedical engineering from Tel-Aviv University, Israel, in 1994; and the Ph.D. degree from the Institute of Sound and Vibration Research (ISVR), Southampton University, U.K., in 1997. At the ISVR, he was appointed Lecturer in 1997 and Senior Lecturer in 2001, working on active control of sound and acoustic signal processing. In 2002, he spent six months as a Visiting Scientist at the Sensory Communication Group, Research Laboratory of Electronics, Massachusetts Institute of Technology (MIT), Cambridge, investigating speech enhancement for hearing aids. He then joined the Department of Electrical and Computer Engineering at Ben-Gurion University as a Senior Lecturer in 2003, and appointed Associate Professor in 2010, and Professor in 2013. He is currently heading the acoustics laboratory, investigating methods for audio signal processing and spatial audio. During 2010-2014 he has served as an associate editor for IEEE Transactions on Audio, Speech and Language Processing, and during 2013-2018 as a member of the IEEE Audio and Acoustic Signal Processing Technical Committee. He also served as an associate editor for IEEE Signal Processing Letters during 2015-2019, IET Signal Processing during 2016-2019, and currently for Acta Acustica. During 2013-2016 he has served as the chair of the Israeli Acoustical Association, and during 2016-2022 as the chair of the Technical Committee on Audio Signal Processing in the European Acoustical Association. He has served as the head of the School of Electrical and Computer Engineering at Ben-Gurion university between 2021-2024. Prof. Rafaely was awarded the British Council’s Clore Foundation Scholarship.
\end{IEEEbiography}

\end{document}